\documentstyle[psfig]{mn}

\newcommand{\lsim}{\raisebox{-3.8pt}{$\;\stackrel{\textstyle <}{\sim}\;$}}

\newcommand{\Msol}{$M_{\odot}$}

\newcommand{\etal}{\mbox{{\rm et~al.\ }}}

\newcommand{\Helium}{\mbox{${\rm {\rm ^{4}He}}$}}
\newcommand{\Carbon}{\mbox{${\rm {\rm ^{12}C}}$}}

\newcommand{\Nitrogen}{\mbox{${\rm {\rm ^{14}N}}$}}

\newcommand{\Oxygen}{\mbox{${\rm {\rm ^{16}O}}$}}

\newcommand{\Magnesium}{\mbox{${\rm {\rm ^{24}Mg}}$}}
\newcommand{\Silicon}{\mbox{${\rm {\rm ^{28}Si}}$}}
\newcommand{\Sulfur}{\mbox{${\rm {\rm ^{32}S}}$}}
\newcommand{\Calcium}{\mbox{${\rm {\rm ^{40}Ca}}$}}
\newcommand{\Fe}{\mbox{${\rm {\rm ^{56}Fe}}$}}

\title{Star formation and chemical evolution in SPH simulations: a statistical
approach}

\author[Lia, Portinari and Carraro]{ Cesario~Lia$^{1,2}$,  
Laura Portinari$^{3,2}$, 
and Giovanni Carraro$^2$ \\
       $^1$ SISSA/ISAS, via Beirut 4, I-34014 Trieste, Italy \\
       $^2$ Dipartimento di Astronomia, Universit\`a di Padova,
 Vicolo dell'Osservatorio 2, I-35122 Padova, Italy \\
       $^3$ Theoretical Astrophysics Center, Juliane Maries Vej 30,
 DK-2100 Copenhagen \O \\
E-mail: {\tt lportina@tac.dk, carraro@pd.astro.it}
}
\date{\tt Submitted: March 2001; accepted: November 2001}
\pubyear{2001}
\begin{document}
\maketitle
\title{Star formation and chemical evolution in SPH simulations}
%
%
\begin{abstract}
In Smoothed Particles Hydrodynamics (SPH) codes with a large number 
of particles, star formation
as well as gas and metal restitution from dying stars can be treated 
statistically. This approach allows 
to include detailed chemical evolution and gas {\mbox{re-ejection}}
with minor
computational effort. Here we report on a new statistical
algorithm for star formation and chemical evolution, especially
conceived for SPH simulations with large numbers of particles,
and for parallel SPH codes.

For the sake of illustration, we present also two astrophysical simulations 
obtained with this
algorithm, implemented into the Tree--SPH code by Lia \& Carraro (2000).\\
In the first one, we follow the formation of  an individual disc--like
galaxy, predict the final structure and metallicity evolution,
and test resolution effects.\\
In the second one we simulate the formation and evolution 
of a cluster of galaxies, to demonstrate the capabilities
of the algorithm in investigating the chemo-dynamical evolution 
of galaxies and of the intergalactic medium in a cosmological context. 
\end{abstract}

\begin{keywords}
hydrodynamics~-~methods:numerical~-~stars: formation ~-~galaxies: evolution~-~
galaxies: chemical evolution
\end{keywords}


\section{Introduction}
The evolution of chemical abundances in the Universe is 
nowadays a subject of utmost importance.
Relevant issues are when and where the first metals were synthetized,
and how they spread to enrich the individual host galaxies, as well as 
the intergalactic (IGM) and intracluster (ICM) medium; see Ferrara 
et~al.\ (2000) and  Chiosi (2000) for recent papers on the subject.

To assess these issues, related to metal production and mixing
into the interstellar medium (ISM), it is necessary to couple chemical 
evolution with hydrodynamical evolution. 
This has been done several times in the past, with different 
techniques. Although in this paper we specifically deal with chemical 
evolution in Smoothed Particle Hydrodynamics (SPH), we like to recall also
some works based on eulerian codes for the hydrodynamics.
The production and distribution of metals over cosmic scales has been 
recently addressed by Cen \& Ostriker (1999) and by Yepes \etal
(1998), who
implemented elementary prescriptions for metal enrichment in their
cosmological simulations. Chemo--dynamical models for individual
galaxies, considering a multi--phase ISM 
have been developed by Burkert \& Hensler (1988), Burkert \etal
(1992), Theis \etal (1992), Samland \etal (1997). 
More recently, 
Recchi \etal (2001) implemented the detailed production and mixing
of many different 
chemical elements in their hydro--code for dwarf 
starburst galaxies.

The first attempt to couple SPH
with chemical evolution is by Steinmetz \& M\"uller (1994), who
study the formation of a Milky Way--like galaxy 
in a cosmological context. Their chemical evolution scheme is rather
simple: they consider the ejection of gas and global metallicity $Z$
from Type~II supernov\ae\ (SN II), occurring over 30 Myr after the
star formation (SF)
episode. These gas and metals produced by a star particle are then 
distributed around over the neighbouring gas particles.

More detailed schemes have been worked out by Raiteri \etal (1996) and
Berczik (1999). Raiteri \etal model the evolution of oxygen and iron,
making use of fitting 
formul\ae\ to follow gas and metal ejection from a star particle over
time. Moreover, they include the delayed iron production by Type~Ia
SN (SN~Ia), by implementing the theoretical rate by Greggio \& Renzini
(1983, hereinafter GR83).
The gas and metals produced by a star particle are distributed around 
over the neighbouring gas particles, as in Steinmetz \& M\"uller (1994). 
Berczik follows the method developed by Raiteri et~al., adding 
planetary nebul\ae, hydrogen 
and helium to the picture. From the chemical point of view, the main
improvement of these models 
is that they avoid the Instantaneous Recycling Approximation 
(IRA, e.g.\ Tinsley 1980) by taking
into account the different stellar sources and 
production timescales of the various elements. 

Carraro \etal (1998) and Buonomo \etal (2000) consider the
evolution of the overall metallicity in the IRA,
with a scheme similar to Steinmetz \& M\"uller (1994). 
However, in the feedback computation they also include the effect of
SN~Ia according to the rate by GR83, and they introduce 
a metal diffusion mechanism instead of the standard SPH smoothing
of metallicity among gas particles.

Recently, Mosconi \etal (2001) presented a new implementation
of chemical evolution in SPH, following the evolution of
very many elements. They consider metal production from SN~II,
essentially instantaneous, and the contribution of SN~Ia. The latter,
rather than following the theoretical rate and time distribution by GR83,
is treated simply as a prompt metal release occurring after a given
time--delay $t_{SNI}$ after the SF episode. The typical time--delay
$t_{SNI}$, 
as well as the number of SN~Ia with respect to SN~II, are introduced 
as free parameters in their simulations.

All the above mentioned models display one or both of the following
numerical drawbacks:

(i) New collisionless star particles are created when SF occurs, while the 
corresponding metal ejecta are redistributed around to other gas
particles. Hence, the number of SF episodes and of ``offspring stars'' 
per gas particle must be artificially damped,
otherwise the overall number of particles gets too large. Also,
the simulation must start with relatively few baryonic
particles, as SF will increase their number substantially in the
course of the simulation;
this hampers the modelling of the early, very interesting phases of
galaxy formation.

(ii) When SF occurs, hybrid particles are created which host both gas and 
different stellar populations. As hybrid particles consist of
both collisional and collisionless sub--components, from the point of
view of the hydrodynamics they introduce artifacts: before
becoming purely collisionless particles, the stars tend to follow for a while
the dynamical evolution of the gas, clearly a spurious effect.

In this paper, we present a new statistical algorithm
of SF and chemical evolution in the context
of N-body SPH simulations, aimed at overcoming the above mentioned
problems. In our approach, in fact, the number of baryonic particles
remains constant throughout the simulation, and particles are either
fully hydrodynamical or fully collisionless. 

Due to its low computational costs,
the new formalism is particularly suited to many--particle
simulations, and the completely local character of the
computation of ``chemical quantities'' makes it convenient   
for parallel codes (Lia 2000, Lia \& Carraro 2000, 
Dalla Vecchia 2001, Lia \etal 2001).

The scheme follows the detailed production of many chemical 
elements over different timescales, avoiding the IRA. In this respect,
we provide updated fitting formul\ae\ for the evolution
of gas and metal release, that can be easily implemented  also
in other particle-based codes. The approach allows to calculate
the abundance evolution of a number
of chemical elements independently, as well as of the global metallicity.
Depending on the objects being modelled, one might in fact be 
interested in monitoring different chemical elements in different cases.

We remark that relaxing the IRA is fundamental not only for 
the sake of chemical evolution (to trace abundance ratios 
of different elements, for instance [$\alpha$/Fe] ratios),
 but also for the effects that delayed
gas restitution may have on the dynamics, as recently discussed by 
Jungwiert \etal (2001). Among others, continuous mass loss from stars 
over a Hubble time contributes 
to solve Robert's time paradox in spiral galaxies (Kennicutt et~al.\ 1994).

The layout of the paper is as follows. In Section~2 we briefly introduce
our TreeSPH code, while in Section~3 we describe the star formation algorithm 
and related feed-back prescriptions.
Section~4 provides a detailed explanation
of our statistical chemical model, whereas Sections~5 to 8 present tests and 
astrophysical applications.
Finally, Section~9 summarizes our findings.


\section{The N-body Tree-SPH code}

The simulations presented here  have been performed using the Tree--SPH
code developed in Carraro \etal (1998), Lia (2000), Lia \& Carraro (2000)
and Dalla Vecchia (2001).
In this code,  the properties of the gaseous  component are
described by means of the SPH technique 
(Lucy 1977; Gingold \& Monaghan 1977), 
whereas the gravitational forces are computed
by means of the hierarchical tree algorithm of Barnes \& Hut (1986), using
the tolerance parameter $\theta=0.8$,   expanding the tree nodes to quadrupole
order, and adopting a spline softening parameter. 
In the SPH method, each  particle represents a fluid element whose
position, velocity, energy, density etc. are followed in  time and space.
The properties of the fluid are locally estimated by an interpolation
which involves the smoothing length $h$.
Convergence tests for the SPH part of the code were performed in 
Carraro \etal (1998).
Finally, cooling tables as a function of the metallicity of the gas have been
taken from Sutherland \& Dopita (1993). 
For further details, we refer the reader to the above mentioned papers.

In this paper we improve upon the SF and Chemical Evolution
sections of our code, as presented here below.


\section{Star Formation and Feed-back}
\label{star_formation}

The formation
of stars is a poorly understood process, and therefore difficult
to model properly. 
The most widely used strategy consists of two steps.

Firstly, an element of fluid must satisfy some conditions
to be  eligible to SF.
The basic criteria adopted in our SPH code
to select the fluid  elements prone to SF  are
(i) the gas particle must be in a convergent flow, and (ii) the gas 
particle must be Jeans unstable.
The conditions are met if the velocity divergence is negative and if the
sound crossing time-scale is shorter than the dynamical time-scale.
The effect of varying the SF criteria 
has been discussed in details
in Buonomo \etal (2000), which the reader is referred to.

Secondly, the fluid element  turns 
into stars according to a suitable star formation rate (SFR),  
which is customarily a reminiscence of the Schmidt (1959) law: 
\[ \frac{d \rho_{\star}}{d t} = - \frac{d \rho_g}{d t} =
c_{\star} \, \frac{\rho_g}{t_{ff}} \]
where $c_\star$ is the star efficency parameter, that we set equal to 0.1,
and $t_{ff}$ is the free-fall timescale.
In our code, this SF law is seen as the probability that a single gas 
particle is completely transformed into a star particle in the  next time step 
(Katz \& Gunn 1991; Katz \etal 1996). This probability is computed as
\begin{equation}
\label{probSF}
P(SF) = 1 - exp (-\frac{c_{\star} \Delta t}{t_{ff}})
\end{equation}
where $\Delta t$ is the particle time step.
The particle is then effectively chosen, or not, to transform 
into a star particle by means of a Monte Carlo method. 
This statistical approach is the more  reliable the larger is 
the number of particles that model the star forming region.

The advantage of this approach is twofold.
First of all, the total number of particles in the simulation 
does not increase,
since a gas particle turns completely into a star particle once it fulfills
all the SF criteria. Besides, baryonic particles are assigned
and always retain all the same mass throughout the simulation. Secondly,  
at increasing number of particles enclosed in a star forming region, 
the probability approach naturally becomes more and more realistic.
Again, this makes it potentially advantageous for algorithms running 
on parallel computers, where large numbers of particles can be managed
(Lia 2000; Dalla Vecchia 2001).

Stars are expected to return energy to the ISM via SN~explosions, and the  
bulk of the energy released in the ISM is known as 
the stellar energy Feed-Back.
In our code, the SN~rates are calculated by the statistical algorithm 
as explained later in this paper, and each SN~ is assumed to contribute 
$10^{50}$~erg of energy to the surrounding medium (Thornton et al 1998).

The key problem here is to know how the released energy is given 
to the ISM, because  the limited resolution of N--body simulations 
does not allow to describe the ISM as a multi-phase medium. 
In Buonomo \etal (2000) we have investigated different prescriptions
for the Energy Feed-Back, suggesting that a good solution is
to release all the energy from SN and other sources to the
thermal budget of the fluid element (Steinmetz \& M\"uller 1994).
As in this scheme the feed-back is localized within the gas particle,
and gets rapidly dissipated by cooling, it turns out to have
no significant effects on the dynamics and/or on the chemical properties
of the gas through the mixing of metals
(which is treated indipendently, see \S~\ref{diffusion}).


\section{The ``statistical'' chemical algorithm}

Once  stars are present, they return to the ISM part of their mass, 
in form of chemically processed gas, and energy, via the stellar
feed--back.
In this paper we extend the probabilistic approach adopted for the SF
process to the energy and chemical feed--back.
Each baryonic particle can be either in the form of gas or in the form of
stars.
The probability that a particle turns from gas to stars is fixed by the
SFR (Eq.~\ref{probSF}). Likewise, each star particle can be assigned 
a certain probability to turn back into gas; and when this takes place,
the particle carries along the corresponding metal production, SN rate, and
energy feed--back.
Notice that in the present approach all baryonic particles, stars and gas,
have the same mass and their total number is conserved. Our statistical 
algorithm is developed
and only valid under this condition. 

In brief, we consider a star particle as a Single Stellar Population (SSP) 
of assigned Initial Mass Function (IMF), and calculate the fraction of 
its mass which turns into re-ejected gas, as a function of its age. 
This mass fraction 
is identified as the 
probability that a star particle of age $t$ becomes a gas particle again 
at time $t + \Delta t$, being $\Delta t$ the particle time step. 
For a Salpeter IMF, for instance, this probability is of the order of 30\%
over a Hubble time, since this is the fraction of the initial mass of the SSP
which is globally re--ejected. By means of a Monte Carlo method, 
at each timestep we transform back into gas a certain fraction of
the star particles; typically a 30\% of the total stars formed 
will have become gas again
by the end of the simulation, after a Hubble time.
Over a large enough number of particles, star formation and
the corresponding gas restitution 
should be well 
represented statistically. Besides, 
this approach 
easily takes into account
that gas restitution from stars is 
diluted in time, with no need to assume the IRA.

Our statistical chemical algorithm is described in detail in the following
sections.

\subsection{Gas restitution}
\label{gasrestitution}

Each star particle is treated as a SSP of total mass $m$ and age $t=T-T_0$,
where $T$ is the present age of the system and $T_0$ is the birth--time
of the star particle.
Ideally, one could follow the detailed chemical, photometric and spectral 
evolution of each SSP by adopting a grid of stellar models 
(e.g.\ Chiosi \etal 1998).
However, in a hydrodynamical code this would heavily increase
the computational load .
To circumvent this, we prefer to approximate 
the amount of released gas, the SN rates and the metal production
by means of analytical formulas, as described here below.

Within a SSP, stellar masses are distributed according to the IMF $\Phi(M)$,
usually a power--law (e.g.\ {\mbox{$\Phi(M) \propto M^{-1.35}$}} 
for a Salpeter IMF). 
A star of given mass $M$ is characterized by a lifetime $\tau(M)$; when it
``dies'',
part of it remains enclosed in a remnant of mass $M_r$ (a white dwarf,
a neutron star or a black hole) while the rest is ejected back to the ISM 
in the form of chemically processed gas. Stellar
models provide $\tau$ and $M_r$ as a function of $M$.
The mass fraction of the SSP which is ejected back in the form of gas by time
$T=T_0+t$ is:
\[ 
E(t) = \int_{M(t)}^{M_u} \frac{M-M_r(M)}{M} \, \Phi(M) \, dM =
\int_{\tau(M_u)}^t e(t') \, dt' \]
where 
\begin{equation}
\label{gasprobab}
e(t) = \left[ \frac{M-M_r(M)}{M} \, \Phi(M) \left( -\frac{dM}{d\tau} \right) 
\right]_{M(t)}
\end{equation}
$M_u = 100$~\Msol\ is the upper stellar mass limit in the IMF,
$M(\tau)$ indicates the mass of a star with lifetime $\tau$, and 
$M(t)=M(T-T_0)$ is thus the smallest stellar mass dying and restituting gas 
by time $T$ (see also Tinsely 1980). For a Salpeter IMF, for instance, 
the global returned gas fraction $E$ over a Hubble time is of the order 
of 30\%.

Accordingly, $e(t)$ of Eq.~(\ref{gasprobab}) is the rate of gas ejection 
in time.
In our statistical approach, this rate is interpreted as the {\it probability}
that a star particle, in time, transforms again into gas. Namely, at each
time step $\Delta t$ the star particle is assigned a probability 
\begin{equation} 
\label{gasprobdt1}
g_t(\Delta t)=\int_t^{t+\Delta t} e(t') \, dt'
\end{equation}
to turn back into gas, and a Monte Carlo method decides whether it does
switch to a gas particle within $\Delta t$. The overall
probability that a star particle ever switches to gas again, integrated
over time, is $E \sim 30$\%; accordingly, of all the star particles
formed at time $T_0$ roughly one third will have returned to be gas
after a Hubble time. Over a sufficiently large number of particles, 
this approach should give 
a fair representation not only of the overall gas restitution, but also of its
rate in time (avoiding the IRA).

\begin{figure}
\psfig{file=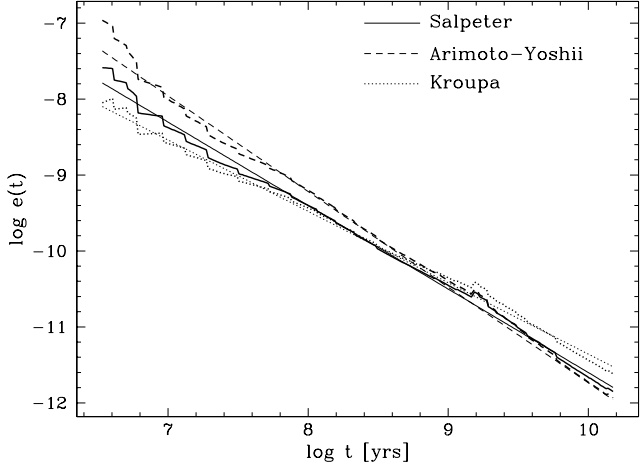,angle=-90,width=8.9truecm}
\caption{Rate of gas restitution for SSPs with different IMFs, as a function
of the age $t$ of the SSP. {\it Thick lines}: numerical results; 
{\it thin lines}: corresponding power--law fits. 
The function $e(t)$
represents as well the probability per year that a star particle
transforms back into a gas particle (see text).}
\label{returnedfig}
\end{figure}

Actually if, say, $N$ particles become star particles at some time $T_0$, 
to yield the correct gas restitution in time the probability~(\ref{gasprobdt1})
should be applied to all of those $N$ particles, at each timestep, 
throughout the simulation.
However, by age $t$ a fraction $E(t)$ of those $N$ particles will have 
already returned to be gas, while only a fraction $1-E(t)$ 
will still be stars. Obviously, the probability to transform back into gas 
at age $t$ can be calculated only for those $N \, [1-E(t)]$ particles that
are still stars at that time, rather than on the base of the whole initial 
population of $N$ particles.
Hence, to recover the correct statistical gas restitution,
the probability~(\ref{gasprobdt1}) must be corrected by a further factor
$[1-E(t)]$, and becomes:
\begin{equation} 
\label{gasprobdt}
g_t(\Delta t)=\frac{\int_t^{t+\Delta t} e(t') \, dt'}{1-E(t)}
\end{equation}
Our test in \S~\ref{singleburst} and Fig.~\ref{gasIMF_sburst} demonstrate
that in this way the correct global gas restitution is statistically recovered
--- which would not be the case if expression~(\ref{gasprobdt1}) 
for the probability were directly used.

We should now determine $e(t)$ as a function of the age $t$ of the SSP.
As mentioned above, in terms of computational effort it is convenient 
to adopt some analytical fit to the numerical results of detailed 
chemical evolution models. We proceed as follows.

\begin{table}
\begin{tabular}{|c|c|c|}
\hline
 Salpeter & Kroupa & Arimoto--Yoshii \\
\hline
 & & \\
~~$e(t)=0.25 \, t^{-1.1}$~~ & ~~$e(t)=0.011 \, t^{-0.94}$~~ & 
~~$e(t)=7.3 \, t^{-1.26}$~~ \\
 & & \\
\multicolumn{3}{l}{$t[yr] > \tau(M_u) = 3.4 \, 10^6$} \\
\hline
\end{tabular}
\caption{Rate of gas restitution $e(t)$ for SSPs with different
IMFs; $e(t)=0$ for $t < 3.4$~Myr, lifetime of the most massive star, 
{\mbox{$M_u = 100$~\Msol}}.
$e(t)$ represents as well the probability per year that a star particle
transforms back into a gas particle (see text).}
\label{returnedtab}
\end{table}

We calculate numerically the function~(\ref{gasprobab})
adopting the stellar lifetimes $\tau(M)$ and remnant masses $M_r(M)$ from 
the Padova evolutionary
tracks (see Portinari \etal 1998, hereinafter PCB98, and references therein). 
Regarding the IMF,
its shape and variations with the environment are still an open 
issue (e.g.\ Scalo 1998); hence we consider three different IMFs suggested
in literature, with the intent of covering a representative range of 
possibilities.
\begin{enumerate}
\item
The standard and widely adopted Salpeter (1955) IMF
\[ \Phi(M) = C_s \, M^{-1.35}~~~~~~~~~~~~C_s=0.1716 \] 
\item
The somewhat steeper IMF by Kroupa (1998), best suited to the Solar 
Neighbourhood:
\[ \begin{array}{l l}
\Phi(M) = &  \left\{
\begin{array}{l l}
C_{k1} \, M^{-0.5}  \, &  \, M < 0.5 \\
C_k \, M^{-1.2}  \, &  \, 0.5 < M < 1 \\
C_k \, M^{-1.7}  \, &  \, M > 1 
\end{array} \right.
\end{array} \]
\[ C_{k1} = 0.48~~~~~~~~~~~~~~~~~~~C_k = 0.295 \]
\item
The  more top--heavy IMF suggested by Arimoto \& Yoshii (1987) for elliptical 
galaxies:
\[ \Phi(M) = C_a M^{-1}~~~~~~~~~~~~C_a = 0.145 \]
\end{enumerate}
$C_s$, $C_a$, $C_k$ and $C_{k1}$ are normalization coefficients fixed so that 
the IMF is 
normalized to unit mass when integrated between the low and high stellar
mass ends (0.1 and 100~\Msol\ respectively). 
With the adopted stellar mass limits, SSPs with the Salpeter or Kroupa IMF
restitute a 30\% of their initial mass in gas, while for SSPs with the more 
top--heavy
IMF by Arimoto \& Yoshii the restitution fraction is larger (about 50\%).

The numerical results for the function~(\ref{gasprobab}) are well fitted by 
power laws (Fig.~\ref{returnedfig}); 
the fitting functions are listed in Table~\ref{returnedtab}. 
In principle, stellar
lifetimes and remnant masses depend also on the metallicity of the star
(PCB98), and hence $e(t)$ also depends on the metallicity
of the parent SSP. However, it is beyond the scope of this paper
to implement chemical evolution in such fine detail; therefore, for gas
restitution (and for the chemical yields below) we adopt fits
independent of metallicity. Notice, however, that
metallicity--dependent prescriptions can in principle be inserted 
in our approach, as a star particle always keeps track of the metallicity
$Z_0$ of its host SSP (as we did for the sole case of the nitrogen yields
in \S~\ref{metalprod}).

With the analytical expressions given for $e(t)$ in Table~\ref{returnedtab}, 
once the IMF is 
chosen it is straightforward to calculate, at each time step, the 
probability~(\ref{gasprobdt}) that a star particle transforms back into gas.
If this happens, in the following discussion we will call it
a ``gas--again particle''.

\begin{table*}
\begin{minipage}{18truecm}
\begin{center}
\begin{small}
\begin{tabular}{|c|c|c|c|c|}
\hline
 & Salpeter & Kroupa & Arimoto--Yoshii & time range [yr] \\
\hline
 & & & & \\
$r_{II}(t)=$ & $2.5 \, 10^{-7} \, t^{-0.43}$ & 
$2.5 \, 10^{-9} \, t^{-0.18}$ & 
$3.2 \, 10^{-5} \, t^{-0.67}$ &
$t \in [3.4 \, 10^6, 7.5 \, 10^7]$ \\
 & & & & \\
\hline
 & & & & \\
$r_{Ia}(t) = \left\{ 
\begin{array}{l}
 \\
 \\
 \\
\end{array} \right.$ & 
$\begin{array}{l}
9.12 \, 10^{-8} \, t^{-0.6} \, 10^{-\frac{8.5 \, 10^{13}}{t^{1.8}}} \\
~~\\
1.74 \, 10^5 \, t^{-1.9} 
\end{array} $ & 
$ \begin{array}{l}
1.15 \, 10^{-8} \, t^{-0.5} \, 10^{-\frac{8.5 \, 10^{13}}{t^{1.8}}} \\
~~\\
1.9 \, 10^5 \, t^{-1.9} 
\end{array} $ & 
$ \begin{array}{l}
1.1 \, 10^{-6} \, t^{-0.7} \, 10^{-\frac{5.3 \, 10^{14}}{t^{1.9}}} \\
~~\\
2.39 \, 10^5 \, t^{-1.9} 
\end{array} $ & 
$ \begin{array}{l}
t \in [7.5 \, 10^7, 2.8 \, 10^9] \\
~~ \\
t > 2.8 \, 10^9 
\end{array} $  \\
 & & & & \\
\hline
\end{tabular}
\end{small}
\end{center}
\end{minipage}
\caption{SN rates for SSPs with different IMFs; the rates of SN II 
and of SN Ia drop to zero out of the respective
time ranges of activity, indicated in the rightmost column.}
\label{rateSNtab}
\end{table*}

\begin{figure}
\psfig{file=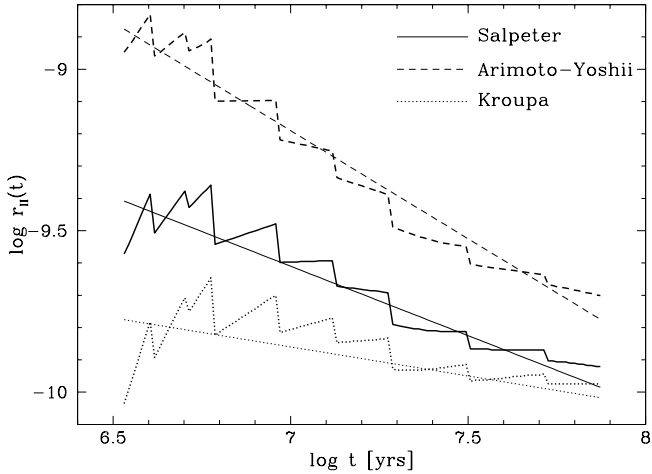,angle=-90,width=9truecm}
\caption{Rate of SN II for SSPs with different IMFs, as a function of the
age $t$ of the SSP. {\it Thick lines}: numerical results; 
{\it thin lines}: corresponding power--law fits.}
\label{rateSNIIfig}
\end{figure}


\subsection{The supernova rate}
\label{SNrates}

The SN rate enters the calculation of chemical evolution and of 
feed-back. We determine the SN rate considering each particle that undergoes
SF as a SSP. We discuss the rates of supernov\ae\ type II 
and type Ia separately.


\subsubsection{The rate of SN II}

SN~II (and SN~Ib,c) originate from single stars of mass 
{\mbox{$M > M_{up}$}}, the minimum stellar mass allowing 
for non--degenerate carbon ignition in the core; we refer to the Padova
stellar tracks, where $M_{up}=6$~\Msol. As $\Phi(M)/M$ is the distribution 
of stellar masses {\it by number} of stars, a SSP of unit total mass 
produces a rate of SN~II in time given by:
\[ \begin{array}{l l}
r_{II}(t) = &  \left\{
\begin{array}{l l}
\left[ \frac{\Phi(M)}{M} \left( -\frac{dM}{d\tau} \right) \right]_{M(t)}
 & ~~\tau(M_u) \leq t \leq \tau(M_{up}) \\
 & \\
0 & ~~otherwise
\end{array} \right.
\end{array} \]
Just as we did for the gas restitution fraction, 
we calculate $r_{II}(t)$ numerically for our three IMFs and fit the
numerical results with suitable analytical functions (Fig.~\ref{rateSNIIfig} 
and Table~\ref{rateSNtab}).

A particle of mass $m$ which underwent SF produces in a time 
step $\Delta t$ a number of SN II given by:
\begin{equation}
\label{numberSNII}
N_{SN II} = m \, \int_t^{t+\Delta t} r_{II}(t') \, dt'
\end{equation}
The factor $m$ must be introduced because $r_{II}(t)$ is the SN rate for 
a SSP of total unit mass (from the normalization of the IMF), 
while our baryonic particles in general have a total mass $m$.


\subsubsection{The rate of SN Ia}

For the rate of SN Ia we follow the scheme of GR83,
assuming SN~Ia to originate from binary systems of total mass $M_B$ between
\[ M_{B,inf} = 3 M_{\odot}~~~~~~~~~{\rm and}~~~~~~~~~~
M_{B,sup} = 2 \, M_{up} = 12 M_{\odot} \]
over timescales set by the lifetime of the secondary star $M_2$ in the system.
In this scenario, the total number of SN~Ia produced by a SSP of age $t$ is:
\[ R_{SN Ia}(t) = A \, \int_{M_{B,inf}}^{M_{B,sup}} \frac{\Phi(M_B)}{M_B} 
\left[ \int_{\mu_{inf}}^{0.5} f(\mu) \, d\mu \right] dM_B \]
where
\[ f(\mu) = 24 \, \mu^2~~~~~~~~~~~~~~~~~~~~~~~~~~ 
\mu = \frac{M_2}{M_B} \in [0, 0.5] \]
\[ \mu_{inf}=\max \left\{ \frac{M_2(t)}{M_B}, \frac{M_B-M_{up}}{M_B} 
\right\} \]
(see GR83, for further details).
$A$ is a parameter
fixed so as to match the observed ratio of SN~II/SN~Ia in Milky Way--type
galaxies; 
for {\mbox{$M_{B,inf}=3$~\Msol}}, $A \sim 0.07$ (e.g.\ PCB98).

Since the timescale of explosion is set by the lifetime of the secondary star,
to calculate the rate of SN Ia in time we first invert the order
of integration:
\[ R_{SN Ia}(t) = A \, \int_{M_2(t)}^{M_{up}} 
		 \left[ \int_{M_{B,min}}^{M_{B,max}} f(\frac{M_2}{M_B}) 
		 \frac{\Phi(M_B)}{M_B^2} dM_B \right] dM_2 \]
where
\[ M_{B,min}=\max \{ M_{B,inf}, \, 2 M_2 \} \]
\[ M_{B,max}=\min \{ M_{B,sup}, \, M_2+M_{up} \} \]
Notice that the formula has been corrected by a factor $M_B^{-1}$ in the inner
integral, with respect to the analogous formula given for a single SF burst
by GR83.

\begin{figure}
\psfig{file=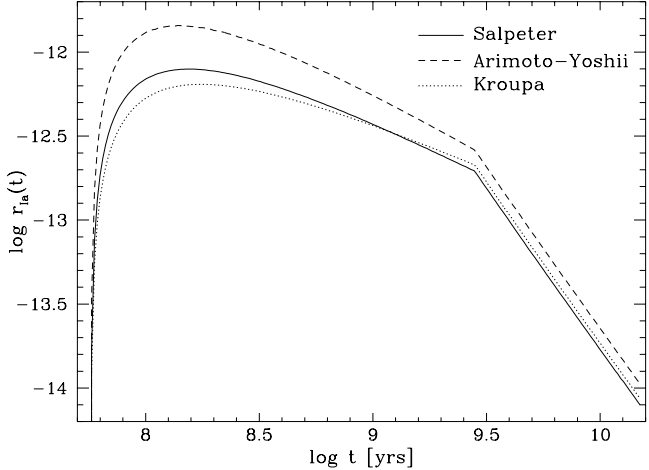,angle=-90,width=9truecm}
\caption{Rate of SN Ia for SSPs of age $t$ and with different IMFs.}
\label{rateSNIfig}
\end{figure}

Accordingly, the rate of SN Ia in time is:
\[ 
r_{Ia}(t) = \left\{
\begin{array}{r}
\medskip
A \, \left[ \left( - \frac{dM_2}{d\tau} \right) \, \int_{M_{B,min}}^{M_{B,max}}
f(\frac{M_2}{M_B}) \frac{\Phi(M_B)}{M_B^2} dM_B  \right]_{M_2(t)}  \\

t \geq \tau(M_{up}) \\
 \\
0~~~~~~~~~~~~~~~~~~~~~~~~~~~~~~~~~~~~~~~~~~~~~otherwise
\end{array} 
\right.
\]
We calculate $r_{Ia}(t)$ numerically for our three IMFs 
and provide analytical fits to the numerical results (Fig.~\ref{rateSNIfig}
and Table~\ref{rateSNtab}).

A particle of mass $m$ which underwent SF produces in a time 
step $\Delta t$ a number of SN~Ia given by:
\begin{equation}
\label{numberSNIa}
N_{SN Ia} = m \, \int_t^{t+\Delta t} r_{Ia}(t') \, dt'
\end{equation}

\subsubsection{The statistical correction}
\label{statcorr}

So far, we have derived the SN rates for a SSP, and consequently 
for a particle which has experienced SF. When we are to calculate the 
feed--back effect that supernov\ae\ have on a given gas particle, 
in principle we need to know the number of SN 
produced in the time step $\Delta t$ by all of its neighbouring 
particles which have experienced
SF at some previous time. These include both star particles and 
gas--again particles.
Although for each of these particles it is straightforward to calculate
the corresponding $N_{SN II}$~(\ref{numberSNII}) and 
$N_{SN Ia}$~(\ref{numberSNIa}), the above mentioned procedure 
would require to determine all the 
neighbouring particles, both gas and stars. On the other hand, for
hydrodynamical purposes only the neighbouring collisional 
(gas and gas--again) particles are
of interest. So, in principle a double calculation of neighbours 
would be necessary, which would require a major computational effort and 
could not be trivially implemented in parallel codes.

\medskip
We thus restrict the calculation of neighbours to gas particles, as usual in
SPH codes. As a consequence, we can determine the contribution to the SN rate
(and to the related feed-back, see Section~4) only 
from neighbouring gas--again particles; 
we should therefore correct for the ``missing'' contribution of star particles.
We proceed as follows.

SN rates within the timestep $\Delta t$ are calculated, rather than for all the
particles that have experienced SF, only for those particles that return to be
gas right {\it within} $\Delta t$. These particles represent a fraction 
$g_t(\Delta t)$ --- given by~(\ref{gasprobdt}) --- of the whole parent 
stellar population.
Accordingly, the ``statistical correction'' must be 
\begin{equation}
\label{corrstat}
 N_{SN} \longrightarrow 
\frac{N_{SN}}{g_t(\Delta t)} 
\end{equation}
Notice that in this approach not all gas--again particles (namely, any
gas particles that have ever experienced SF in their past) contribute
to the SN rates, but only those that become gas--again ``right now''.
This ensures that the SN explosions (and the metal production,
see \S~\ref{metalprod} below) take place exactly where the parent stellar 
component is located.

The statistical correction~(\ref{corrstat}) is quite large: a whole stellar 
population
is sampled by the fraction of it which dies within the timestep $\Delta t$.
This fraction can become very small, especially for small timesteps and/or 
at advanced ages after the SF episode (see the behaviour of $e(t)$ in time,
Fig.~\ref{returnedfig}). 
At late times, the hydrodynamical timestep of the system is 
very small with respect to the rate of gas restitution, hence statistical
fluctuations in the number of old star particles turning to gas--again tend
to become quite large. This might induce large fluctuations in the effective
release of SN and metals with respect the smooth theoretical time evolution
of a SSP. To hamper these fluctuation, the probability 
that a star particle becomes
gas--again is computed, rather than over the dynamical timestep,
over a ``chemical timestep'' which increases in proportion 
with the age of the SSP, so as
to smoothen the related  increase of the statistical noise.
For the sake of clarity, from here on we indicate by $\Delta t$ 
and $\delta t$ the chemical and dynamical timestep respectively.
The chemical timestep $\Delta t$ is chosen to be the minimum
between $\delta t$ and one tenth of the age $t$ of the SSP.
We also put an upper limit to $\Delta t$ of $2 \times 10^{8}$~yr, so that
it does not become excessively large from the dynamical point of view,
when we are dealing with stars that are a few Gyr old. $2 \times 10^{8}$~yr
corresponds, for instance, to the typical revolution period of the Sun in the
Galaxy.

In detail, our algorithm works as follows. For a star particle, the ``Monte
Carlo chance'' to transform to gas--again is activated once every $\Delta t$,
rather than at each dynamical step $\delta t$, with a corresponding
probability~(\ref{gasprobdt}) also calculated for $\Delta t$. If the
star particle is selected to become a gas--again particle, it ``produces''
a number of SN~(\ref{corrstat}) also calculated on the base of $\Delta t$.
Before the next dynamical timestep, this particle is added 
to the list of gas particles for the calculation of neighbours, 
and from then on it behaves just like any other gas particle.
Finally, it is worth noting that 
the energy feed-back is released over the hydrodynamical
timestep, and the energy balance between cooling and heating is 
calculated over the  hydrodynamical timestep, as well.

The various test applications presented later in this paper show
that this statistical approach, with the adopted chemical
timestep, yields sensible results.


\subsection{Chemical enrichment}

The calculation of the chemical enrichment of the gas proceeds in two steps:
\begin{enumerate}
\item
calculation of metal production by SSPs in particles 
which have experienced SF;
\item
metal diffusion among gas particles.
\end{enumerate}
In the tables we provide the necessary information to calculate the production
of various chemical elements, as well as that of the overall metallicity,
independently of one another. In fact, according to the particular object
being modelled, one might be interested in tracking different chemical 
elements because available abundance data also depend on the
class of object considered. Hence, in each simulation one can select and 
``switch on'' the elements of interest. The hydrogen abundance, 
necessary to express chemical abundances in the usual [element/H] dex scale,
can be obtained for each particle as:
\[ X = 1-Y-Z \]
where $Y$ represent the helium mass fraction and $Z$ the overall metallicity.

For the sake of clarity, here below we will describe our implementation of 
chemical evolution in terms of a single ``metal parameter'' $Z$. This parameter
is meant to represent the chemical abundance of any specific chemical element
 --- helium or metals.

\begin{table*}
\begin{minipage}{18truecm}
\begin{center}
\begin{tabular}{|r r|c|c|c|c|}
\hline
\multicolumn{2}{|c|}{chemical element} & Salpeter & Kroupa & Arimoto--Yoshii & 
time range  [yr] \\
\hline
\hline
$Z$ & $p_Z(t)=$ & $453 \, t^{-1.7}$ & $10.5 \, t^{-1.5}$ & $5919 \, t^{-1.8}$ &
$t \in [3.4 \, 10^6, 15 \, 10^9]$ \\
\hline
\Helium & $p_{He}(t)=$ & $2.8 \, t^{-1.37}$ & $0.14 \, t^{-1.21}$ & 
$80 \, t^{-1.52}$ & $t \in [3.4 \, 10^6, 15 \, 10^9]$ \\
\hline
\Oxygen & $p_O(t) = \left\{ 
\begin{array}{l}
 \\
 \\
\end{array} \right.$ & 
$\begin{array}{l}
7.3e{-10} \\
7.3e4 \, t^{-2} 
\end{array} $ & 
$\begin{array}{l}
4.2e{-10} \\
4.2e4 \, t^{-2} 
\end{array} $ & 
$\begin{array}{l}
1.8e{-9} \\
1.8e5 \, t^{-2} 
\end{array} $ & 
$ \begin{array}{l}
t \in [3.4 \, 10^6, 10^7] \\
t \in [10^7, 3.4 \, 10^7]
\end{array} $  \\
\hline
\Fe & $p_{Fe}(t) = \left\{ 
\begin{array}{l}
 \\
 \\
\end{array} \right.$ & 
$\begin{array}{l}
5.1e{-11} \\
5.1e3 \, t^{-2} 
\end{array} $ & 
$\begin{array}{l}
3.3e{-11} \\
3.3e3 \, t^{-2} 
\end{array} $ & 
$\begin{array}{l}
 1.1e{-10} \\
 1.1e4 \, t^{-2} 
\end{array} $ & 
$ \begin{array}{l}
t \in [3.4 \, 10^6, 10^7] \\
t \in [10^7, 3.4 \, 10^7]
\end{array} $  \\
\hline
\Magnesium & $p_{Mg}(t) = \left\{ 
\begin{array}{l}
 \\
 \\
\end{array} \right.$ & 
$\begin{array}{l}
4.6e{-11} \\
4.6e3 \, t^{-2} 
\end{array} $ & 
$\begin{array}{l}
2.5e{-11} \\
2.5e3 \, t^{-2} 
\end{array} $ & 
$\begin{array}{l}
1.2e{-10} \\
1.2e4 \, t^{-2} 
\end{array} $ & 
$ \begin{array}{l}
t \in [3.4 \, 10^6, 10^7] \\
t \in [10^7, 3.4 \, 10^7]
\end{array} $  \\
\hline
\Silicon & $p_{Si}(t) = \left\{ 
\begin{array}{l}
 \\
 \\
\end{array} \right.$ & 
$\begin{array}{l}
5.8e{-11} \\
5.8e3 \, t^{-2} 
\end{array} $ & 
$\begin{array}{l}
3.6e{-11} \\
3.6e3 \, t^{-2} 
\end{array} $ & 
$\begin{array}{l}
1.3e{-10} \\
1.3e4 \, t^{-2} 
\end{array} $ & 
$ \begin{array}{l}
t \in [3.4 \, 10^6, 10^7] \\
t \in [10^7, 3.4 \, 10^7]
\end{array} $  \\
\hline
\Sulfur & $p_S(t) = \left\{ 
\begin{array}{l}
 \\
 \\
\end{array} \right.$ & 
$\begin{array}{l}
2.9e{-11} \\
2.9e3 \, t^{-2} 
\end{array} $ & 
$\begin{array}{l}
1.8e{-11} \\
1.8e3 \, t^{-2} 
\end{array} $ & 
$\begin{array}{l}
6.6e{-11} \\
6.6e3 \, t^{-2} 
\end{array} $ & 
$ \begin{array}{l}
t \in [3.4 \, 10^6, 10^7] \\
t \in [10^7, 3.4 \, 10^7]
\end{array} $  \\
\hline
\Calcium & $p_{Ca}(t) = \left\{ 
\begin{array}{l}
 \\
 \\
\end{array} \right.$ & 
$\begin{array}{l}
4.2e{-12} \\
4.2e2 \, t^{-2} 
\end{array} $ & 
$\begin{array}{l}
2.7e{-12} \\
2.7e2 \, t^{-2} 
\end{array} $ & 
$\begin{array}{l}
9.1e{-12} \\
9.1e2 \, t^{-2} 
\end{array} $ & 
$ \begin{array}{l}
t \in [3.4 \, 10^6, 10^7] \\
t \in [10^7, 3.4 \, 10^7]
\end{array} $  \\
\hline
\Carbon & $p_C(t) = \left\{ 
\begin{array}{l}
 \\
 \\
 \\
\end{array} \right.$ & 
$\begin{array}{c}
1.2e{-10} \\
1.2e4 \, t^{-2} \\
1.1e{-6} \, t^{-0.7} - 3e{-13}
\end{array} $ & 
$\begin{array}{c}
5.8e{-11} \\
5.8e3 \, t^{-2} \\
2.9e{-8} \, t^{-0.5} - 8e{-13}
\end{array} $ & 
$\begin{array}{c}
3.6e{-10} \\
3.6e4 \, t^{-2} \\
6.4e{-5} \, t^{-0.9} - 2e{-13}
\end{array} $ & 
$ \begin{array}{l}
t \in [3.4 \, 10^6, 10^7] \\
t \in [10^7, 3.4 \, 10^7] \\
t \in [2 \, 10^8, 5 \, 10^9] \\
\end{array} $  \\
\hline
\Nitrogen & $p_{Ns}(t)=$ & $7.7 \, Z_0 \, t^{-1.4}$ & 
$0.98 \, Z_0 \, t^{-1.3}$ & $520 \, Z_0 \, t^{-1.6}$ & 
$t \in [3.4 \, 10^6, 15 \, 10^9]$ \\
 & & & & & \\
\multicolumn{2}{|c|}{\footnotesize
$p_{Np}(t) = \left\{ 
\begin{array}{l}
 Z_0 < 0.004 \\
 Z_0 \in [0.004,0.02] \\
 Z_0 > 0.02 \\
\end{array} \right.$} & 
$\begin{array}{l}
3.3e{-12} \\
\frac{3.3e{-18}}{Z_0^{5/2}} \\
5.8e{-14}
\end{array} $ & 
$\begin{array}{l}
3.3e{-12} \\
\frac{3.3e{-18}}{Z_0^{5/2}} \\
5.8e{-14}
\end{array} $ & 
$\begin{array}{l}
4.6e{-12} \\
\frac{4.7e{-18}}{Z_0^{5/2}} \\
8.3e{-14}
\end{array} $ & 
$ \begin{array}{l}
 \\
t \in [10^8, 2.5 \, 10^8] \\
 \\
\end{array} $  \\
\hline
\end{tabular}
\end{center}
\end{minipage}
\caption{Rate of release of chemical yields of various elements for SSPs 
of age $t$ and with different IMFs; the rates drop to zero out of the 
respective time ranges, indicated in the rightmost column. 
In the case of nitrogen, we distinguish the secondary and the primary 
components and give metallicity--dependent prescriptions; $Z_0$ is the initial
metallicity of the SSP. 
Needless to say, the total yield for nitrogen is $p_N(t)=p_{Ns}(t)+p_{Np}(t)$
(see text for details).}
\label{pZtab}
\end{table*}


\subsubsection{Metal production of a star particle}
\label{metalprod}

Let's consider a gas particle of metallicity $Z_0$ which becomes a star 
particle at time $T_0$: it effectively hosts a SSP composed of a distribution 
of stellar masses 
$\Phi(M)$, all born at time $T_0$ out of gas with homogeneous metallicity
$Z_0$.
As time progresses, stars of smaller and smaller mass $M$ die,
each ejecting a mass of metals given by:
\[ M_Z = y_Z + Z_0 (M-M_r) \]
The first term $y_Z$ is the stellar yield, i.e.\ the newly synthesized 
metals (see the definition by Tinsley 1980 as revised by Maeder 1992);
the second term is the metals present in the star from birth and re-ejected.
The mass fraction that a SSP releases in the form of metals up to age $t$ is:
\[ \begin{array}{l l}
E_Z(t) & = \int_{M(t)}^{M_u} \frac{M_Z}{M} \, \Phi(M) \, dM \\
 & \\
       & = \int_{M(t)}^{M_u} \frac{y_Z}{M} \, \Phi(M) \, dM \,+\, 
           Z_0 \, \int_{M(t)}^{M_u} \frac{M-M_r}{M} \, \Phi(M) \, dM \\
& \\
       & = \int_{\tau(M_u)}^t \left[ y_Z \frac{\Phi(M)}{M} 
\left( - \frac{dM}{d\tau} \right) \right]_{M(t')} dt'  \,+\, Z_0 \, E(t)
\end{array} \]
Evidently, summing the $E_Z$'s over all the chemical elements one expects:
\[ \sum E_Z (t) = E(t) \]
The rate of metal ejection becomes:
\[ e_Z(t) = p_Z(t) \,+\, Z_0 \, e(t) \]
which is determined once we know
\begin{equation}
\label{pZfunc}
 p_Z(t) = \left[ y_Z \, \frac{\Phi(M)}{M} \left( - \frac{dM}{d\tau} \right) 
\right]_{M(t)}
\end{equation}
for the chemical element $Z$ of interest, since $e(t)$ has already been 
calculated in \S~\ref{gasrestitution}.

To calculate $p_Z(t)$, we adopt the stellar yields $y_Z$
by PCB98 for the case of massive stars, and by Marigo 
(2001, her case $\alpha=1.68$) for low and intermediate mass stars. 
Similarly to what we did for the returned gas fraction and for the SN rates, 
we calculate the functions~(\ref{pZfunc}) numerically, for 
several chemical elements and for the three IMF cases.
Then we provide suitable analytical fits to the numerical results, to be used
in the hydrodynamical code. Our fitting functions are listed in 
Table~\ref{pZtab}. As in the case of the returned gas fraction, we neglect 
in general the dependence of stellar yields on the initial
metallicity of the SSP.

The derived fitting functions reflect the different nucleosynthetic
history of the various elements.
Both helium and metals (meant as the overall metallicity) are expelled
over the whole mass range; although the bulk comes from massive stars,
the contribution from stars of intermediate and low mass cannot be neglected. 
Therefore,
the ``onset'' of the fitting functions for $He$ and $Z$ corresponds to the
lifetime of the most massive star (3.4~Myr) and the production continues 
forever;
calculations have though been stopped at 15~Gyr, beyond that (which
is beyond the age of the Universe anyways) the production
can be considered negligible. The production rates rapidly decrease in time,
due to the behaviour of stellar lifetimes with mass (cf.\
the factor $dM/d\tau$ in Eq.~\ref{pZfunc}). The Arimoto--Yoshii IMF case
is the most skewed toward massive stars, hence the metal production is the
largest 
and it presents the steepest drop with time as it is
the most dominated by massive stars. The global metal production
is lower and its decrease with time is slower going to the Salpeter 
and then to the Kroupa IMF case.

Oxygen production, due to massive stars,
is limited to the range of lifetimes of SN II progenitors
(3.4 to 34~Myr). Especially for the most massive stars, 
oxygen production is sensitive to metallicity (PCB98). 
In the framework of a chemical
algorithm handy enough to be implemented in hydrodynamical codes,
it is not worth entering so much detail and we rather give
average estimates of oxygen yields. 
Again, the global
oxygen production is largest for the Arimoto--Yoshii IMF case and decreases
going to the Salpeter and to the Kroupa case. 

Similar trends, and comments, apply to the production of iron, magnesium,
silicon, sulfur and calcium, also ejected by massive stars.
Notice that we are discussing here only chemical yields from single 
stars; the contribution of SN~Ia (binaries) will be added later. 
Theoretical magnesium yields are known to be underestimated with respect
to observations (Timmes \etal 1995, Thomas \etal 1998, PCB98,
Chiappini \etal 1999), while iron yields are sometimes suggested to be 
a bit high (though still within the intrinsic uncertainty of a factor of 2
in SN models, Timmes \etal 1995). The latter point is true especially when
considering the latest estimates of the solar abundance 
[Fe/H]$_{\odot}$=7.5, 
rather than the old value of 7.67. In Table~\ref{pZtab},
entries for magnesium and iron have been optimized on the base of 
observational indications for the Solar Vicinity.

The nucleosynthetic history of carbon is quite composite. 
Carbon is ejected by massive stars through stellar winds and SN II
explosions, while intermediate and low--mass stars contribute to its
production in their {\mbox{TP-AGB}} phase. In both cases,
theoretical carbon yields are sensitive to metallicity effects
(PCB98, Marigo 2001). From the point of view of observations, the role of
the various contributors and of metallicity dependence is still debated
({\mbox{Prantzos}} \etal 1994, Gustafsson \etal 1999, Garnett \etal 1999, 
Henry \etal 2000, Carigi 2000, Liang et~al.\ 2001). 
Considering these uncertanties,
for the purpose of the present chemical algorithm we simply provide 
carbon yields averaged over metallicity
both for massive stars (time range 3.4--34~Myr) and for the delayed
contribution of lower mass stars (0.2--5~Gyr). 
Also for carbon 
the entries in Table~\ref{pZtab} have been adjusted on the base of
observational constraints for the Solar Vicinity.

The stellar nucleosynthesis of nitrogen is also complex, and the 
distinction between its primary and secondary production (Tinsley 1980) 
is of prime importance to interpret observational
evidence (e.g.\ Larsen \etal 2001 and references therein).
For nitrogen, therefore, we consider it worth treating 
the secondary and primary components separately ($p_{Ns}$ and $p_{Np}$), 
with metallicity dependent prescriptions.
The secondary component is, by definition,
directly proportional to the metallicity of the
parent SSP 
and is produced by stars of all masses
(from 3.4~Myr to 15~Gyr).
Primary production occurs in intermediate mass stars of 3.5--5~\Msol\
(time range $\sim$100--250~Myr) and is very sensitive to metallicity, 
being more efficient at
lower metallicities (Marigo 2001, her case $\alpha$=1.68 adopted here). 
Hence we give a metallicity dependent
fit also for the primary component. 
For carbon and nitrogen, the relative importance of the production by
intermediate and low mass stars with respect to that by massive stars
increases when moving from the Arimoto--Yoshii to the Salpeter to the Kroupa
IMF, due to the corresponding shift of the SSP towards 
less massive stars.

\medskip\noindent
\underline{The contribution of SN Ia}

\noindent
Table~\ref{pZtab} provides the rate of metal release in time from single stars
in a SSP of total unit mass. We should then add the contribution of SN~Ia,
which originate in binary systems.
This is easily done since we know the
rate of SN~Ia in time (Table~\ref{rateSNtab}); for each SN Ia, we adopt
the chemical ejecta $M_Z^{Ia}$ from the W7 model by Iwamoto \etal (1999). 
The total metal
release by the SSP, including the contribution of SN Ia, is given by:
\[ p_Z^{tot}(t) = p_Z(t) + M_Z^{Ia} \times r_{Ia}(t) \]
as listed in Table~\ref{pZtottab}.

\medskip\noindent
Hence, within a timestep $\Delta t$ a SSP of age $t$ 
releases an amount of metals given by:
\begin{footnotesize}
\[ \Delta m_Z = \int_t^{t+\Delta t} e_Z(t') dt' =
\int_t^{t+\Delta t} p_Z^{tot}(t') \, dt' + Z_0
\int_t^{t+\Delta t} e(t') dt' \]
\end{footnotesize}
or, equivalently, ejects an amount $\int_t^{t+\Delta t} e(t') \, dt'$
of gas with metallicity:
\begin{equation}
\label{ZtSSP}
Z_t(\Delta t) = \frac{\Delta m_Z}{\int_t^{t+\Delta t} e(t') \, dt'} 
\end{equation}
%

\begin{table}
\begin{center}
\begin{tabular}{c l}
\hline

\medskip
\Carbon & $p_C^{tot}(t) = p_C(t) + 4.83 \, 10^{-2} \, r_{Ia}(t)$ \\

\medskip
\Nitrogen & $p_N^{tot}(t) = p_N(t) + 1.16 \, 10^{-6} \, r_{Ia}(t)$ \\

\medskip
\Oxygen & $p_O^{tot}(t) = p_O(t) + 0.143 \,\,\, r_{Ia}(t)$ \\

\medskip
\Magnesium & $p_{Mg}^{tot}(t) = p_{Mg}(t) + 8.5 \, 10^{-3} \, r_{Ia}(t)$ \\

\medskip
\Silicon & $p_{Si}^{tot}(t) = p_{Si}(t) + 0.154 \,\,\, r_{Ia}(t)$ \\

\medskip
\Sulfur & $p_S^{tot}(t) = p_S(t) + 8.46 \, 10^{-2} \, r_{Ia}(t)$ \\

\medskip
\Calcium & $p_{Ca}^{tot}(t) = p_{Ca}(t) + 1.19 \, 10^{-2} \, r_{Ia}(t)$ \\

\medskip
\Fe & $p_{Fe}^{tot}(t) = p_{Fe}(t) + 0.626 \,\,\, r_{Ia}(t)$ \\

\medskip
$Z$ & $p_Z^{tot}(t) = p_Z(t) + 1.4 \,\,\, r_{Ia}(t)$ \\
\hline
\end{tabular}
\end{center}
\caption{Total rate of release of chemical yields from a SSP, including
the contribution of SN Ia. $r_{Ia}(t)$ is given in 
Table~\protect{\ref{rateSNtab}} }
\label{pZtottab}
\end{table}


\subsubsection{Metal release from the star particle}

\noindent
A particle which has experienced SF is considered to host a SSP, which 
releases gas and metals in time as calculated above.
From the chemical point of view, that is, within such 
a particle there are ``hidden'' stellar and gaseous sub-components evolving 
in time. From the point of
view of hydrodynamics, however, in our simulations we do not resolve these 
sub-components: a baryonic particle is labelled either as gas or as star, 
and behaves accordingly either as collisional or collisionless matter.
We must hence translate the chemical enrichment of a SSP calculated above into 
chemical enrichment of the gas particles. 

Just as in the case 
of gas restitution and of the SN rates (\S\ref{gasrestitution} and 
\S\ref{SNrates}), we resort to a statistical approach. A particle which 
has experienced SF may either remain a star forever, or become at some point
a gas--again particle. As long as it remains a star particle, it is assigned 
the metallicity $Z_0$ of its SSP; when at some timestep $\Delta t$ 
it turns into a gas--again particle, it is assigned the composition 
$Z_t(\Delta t)$ of the gas released by the SSP within $\Delta t$, given 
by~(\ref{ZtSSP}).

Once more, the main limit of this approach may reside in poor statistics:
the metal production of a whole stellar population is sampled by the, 
possibly small, fraction of it which dies within the timestep $\Delta t$.
A large enough number of particles is necessary to obtain
a meaningful statistical sampling, and the ``chemical timestep'' 
must be chosen appropriately, generally longer that the hydro--dynamical
timestep (see \S\ref{statcorr}). Our test for the single burst case
and our simulations below indicate that our algorithm works already
with a few thousand baryonic particles.


\subsection{Metal diffusion in the gas}
\label{diffusion}
Finally, we must describe the chemical evolution of the overall population
of gas particles, both gas--again particles and those which have never been 
stars, which proceeds by metal diffusion. Particles which were never stars
can acquire metals from neighbouring gas--again particles through diffusion.
Conversely, a gas--again particle is initially assigned the metallicity 
$Z_t(\Delta t)$ from Eq.~(\ref{ZtSSP}) as discussed above; from then on, 
its composition 
will evolve only by metal exchange with other nearby gas particles,
with no further memory that it had been a star in the past.
For the sake of metal diffusion, that is, once the initial metallicity 
of a gas--again particle has been assigned 
there is no need to distinguish between gas and gas--again particles.

At each time step, 
we diffuse metallicity among gas particles according to the
scheme originally developed by Groom (1997), which consists
of an SPH translation of the usual diffusion equation:
\begin{equation}
\frac{dZ}{dt} = - \kappa \nabla^{2} Z
\end{equation}
\noindent
where  $\kappa$ is the diffusion coefficient.
In principle, this scheme is
more physically grounded
than the 
widely used 
SPH-smoothing, in 
which metals are SPH-spread among the neighbouring
particles. It relies on the idea
that the diffusion of metals is driven by
SN~explosion and energy injection in the interstellar medium
during the SN~remnant phase.

We derive the 
diffusion coefficient from the models
by Thornton \etal (1998). The typical size of a SN remnant 
after $10^{6}$ yr from the explosion is
50-100 pc, whereas the typical velocity of the gas accelerated by the SN
remnant at that time is 40-60 km/sec. In this way we obtain:
\[
\kappa = (50~\rm km~sec^{-1})(1.85 \times 10^{15} km) \, km^{2} \, sec^{-1}
\]
\noindent
The SPH translation, which requires second derivatives of the kernel,
is
\begin{displaymath} 
\frac{dZ_{i}}{dt} = - \sum^{N}_{j=1} \kappa m_{j} \left(
\frac{2}{\rho_{j}+
\rho_{i}}\right) \left( Z_{i} - Z_{j} \right) \times
\end{displaymath} 
\begin{equation}
~~~~~~~~~~~~~~~ | \nabla^{2}W_{ij}(|\vec{r}_{i}-\vec{r}_{j}|,h_{ij})|
\end{equation}
This SPH algorithm was successfully tested by Groom (1997) versus
suitable analytical exact counterparts.

A possible drawback of the adopted diffusion scheme could be   
the use of the $2^{nd}$ derivative of the smoothing function, which
may introduce spurious fluctuations in the resulting metallicity.
The use of a spline kernel, which is continuous with its 1$^{st}$ and
2$^{nd}$ order derivatives, ensures that unphysical fluctuations
in the diffused quantities (metallicity in our case) do not appear.

\begin{figure}
\psfig{file=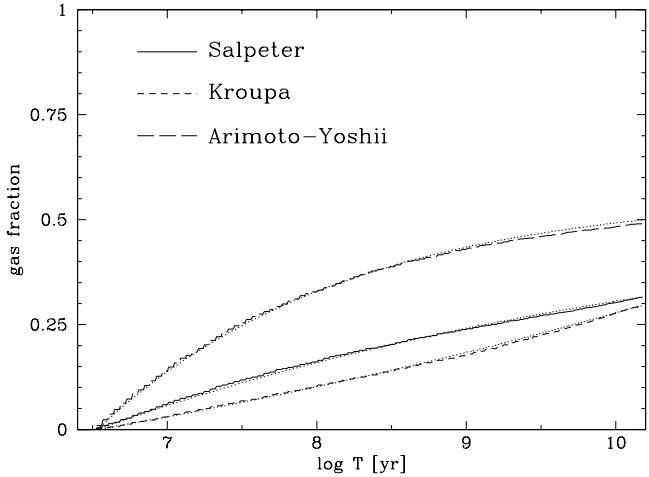,width=9truecm,angle=-90}
\caption{Returned gas fraction in time for a single burst of star formation 
involving 5000 particles. Thin dotted lines represent the exact analytical
expectations.}
\label{gasIMF_sburst}
\end{figure}

\begin{figure*}
\centerline{\psfig{file=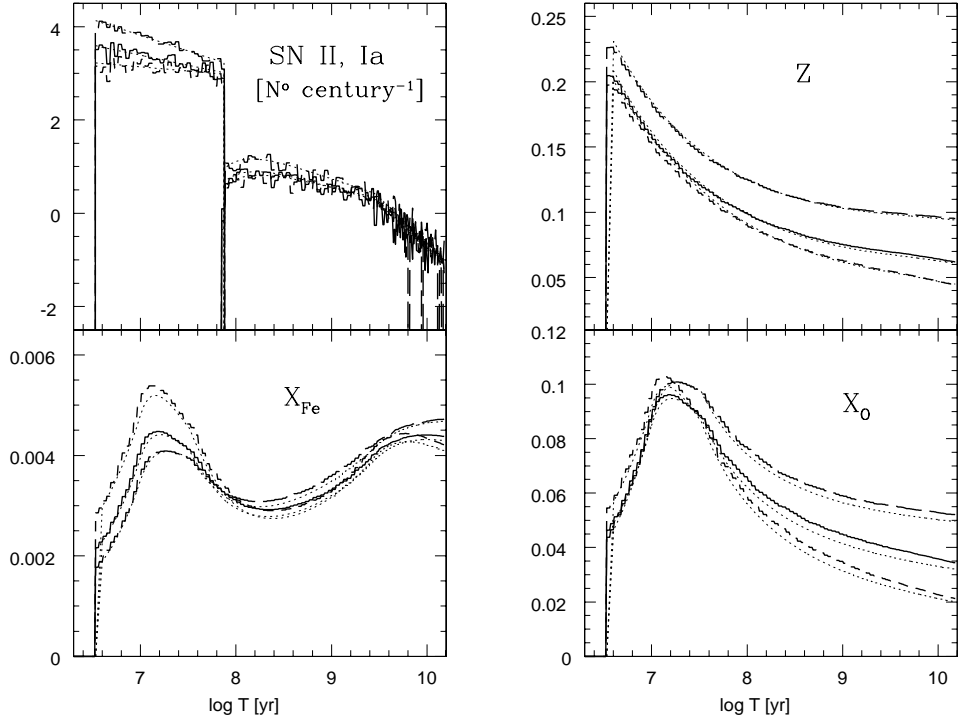,width=14cm,angle=-90}}
\caption{Chemical evolution for a single burst of star formation involving 
10$^{11}$~\Msol. {\it Upper-left panel}: cumulative rates of SN~II 
and~Ia ($T <$ and $> 7.5e7$~yr, respectively)  
in logarithm of the number of events per century.
{\it Upper-right panel}: evolution of the total metallicity.
{\it Lower panels}: evolution of iron (on the left) and oxygen (on the right) 
abundances.
{\it Solid line}: numerical results for the Salpeter IMF; 
{\it short--dashedline}: Kroupa IMF; {\it long--dashed line}: 
Arimoto--Yoshii IMF. The {\it dotted lines} are the corresponding 
exact analytical predictions.}
\label{singleburstIMF}
\end{figure*}


\section{A single burst}
\label{singleburst}
A first useful test for our algorithm is to model the evolution of the various
chemical quantities (gas fraction, metallicity etc.) in the simple case 
of a single burst of SF. This test focuses entirely 
on the implemented chemical network, uncoupled in this case from 
hydrodynamical effects. It allows us on one hand to visualize the expected 
chemical production of a SSP, and on the other hand to check 
the validity of the statistical approach.

The single burst model is realized 
with 5000 particles of gas, 
of total mass of $10^{11}~M_{\odot}$, 
turned into stars instantaneously at the beginning of the simulation
($T_0=0$, $Z_0=0$). These star particles are let evolve afterwards according 
to the statistical prescriptions for
gas and metal restitution given above, without any further 
SF episodes.

It is sufficient to model the evolution of a couple of characteristic elements
with different production timescales, to assess the capability of the
statistical approach in describing detailed chemical evolution in time.
We choose to follow the evolution of oxygen (representative of
$\alpha$--elements produced by massive stars over short time scales), of
iron (representative of elements with a delayed contribution), 
and of the global metallicity.

As everything in our algorithm (SN rates, metal production etc.) is entirely
governed by the statistical gas restitution, it is fundamental to check this
quantity first of all. Fig.~\ref{gasIMF_sburst} shows the evolution of the 
returned gas fraction in time for our single burst test, for the three IMF 
cases, as compared to the exact expectations computed directly
from the analytical fitting functions. The response of the algorithm is very
good already with $\sim$5000 particles.

In Fig.~\ref{singleburstIMF} we display the results for SN rates and for
the evolution of metallicity, iron and oxygen abundances in the returned gas,
again compared to the analytical predictions.

The results reproduce the expected trends for a SSP, and although
statistical fluctuations are apparent, for instance, in the predicted SN rates,
the overall behaviour is traced quite well. Metallicity is very high 
at the beginning, when the highly metal enriched gas from massive stars 
is expelled.
Later, this extremely metal rich gas is diluted by more metal poor gas
ejected by long--lived stars, so that metallicity decreases. The oxygen 
abundance
follows the same trend, and represents roughly a half of the global 
metallicity, being oxygen in fact the most abundant of all metals. 
The iron abundance also decreases after the initial peak due to massive
stars, but contrary to
oxygen it later stabilizes and even increases again thanks to the 
contribution of SN~Ia.
The oxygen--to--iron ratio correspondingly changes from super-solar
values at early times, when SN II dominate the chemical enrichment, to
solar and sub-solar values at later times.

The absolute number of SN is largest for the Arimoto--Yoshii IMF, which is
the most top--heavy, and decreases moving to the Salpeter and to the Kroupa
case; the same trend is seen in the respective metallicities and oxygen
abundances, dominated by the production of massive stars.
The {\it relative} number of SN Ia to SN II, however, decreases when going 
from Arimoto--Yoshii to Salpeter to Kroupa because these latter IMFs are more 
skewed toward smaller stellar masses. Hence, the final [O/Fe] ratio typical 
of the Arimoto--Yoshii SSP is larger than for the other IMF cases.

Within the range of massive stars ($T \lsim 10^8$~yr), one can notice that 
the typical iron abundance in the ejected gas increases going from the 
Arimoto-Yoshii to the Salpeter to the Kroupa IMF. This is due to the fact
that SN from progenitors in the low mass end of massive stars (say, 8-15~\Msol)
are expected to produce a higher amount of iron, relative to the global
mass ejection or to the oxygen production, than stars in the higher mass
range. Hence, the Arimoto-Yoshii IMF favours the highest masses with
very oxygen--rich, and less iron--rich, ejecta;
on the contrary, the
Kroupa IMF is more skewed toward SN with a higher relative iron production.
As a consequence, the typical [O/Fe] ratio of SN~II ejecta from an 
Arimoto-Yoshii
SSP is higher than from a Salpeter SSP than for a Kroupa SSP. Typical values 
are [O/Fe]=+0.56 for Arimoto-Yoshii, +0.50 for Salpeter, +0.45 for Kroupa.
The theoretical estimates of iron production as a function of mass, on the 
other hand, are very uncertain and this detailed mass effect might be somewhat
spurious. However, all of these values are compatible with empirical evidence
for the $\alpha$-enhanced stellar population of the Galactic halo (see e.g.\
the data by Carretta \etal 2000 in Fig.~\ref{discOsuFe}).

The Kroupa IMF has been adopted for the simulations presented in the following.

\section{A disc--like galaxy}
As a first astrophysical test--application of our code we run 
a simulation for an individual galaxy.

The initial configuration for this object is a
a spherical DM halo with a density profile $\propto 1/r$ (see Lia \etal 2000
for a justification of this choice).
The total mass of the system is $1.3 \, 10^{12}$~\Msol\ with a baryonic 
fraction equal to  $0.1$.  The galaxy is modeled using 30,000 baryonic
particles and 15,000 DM particles.
Accordingly, the mass of a DM particle is $7.8 \times 10^{7} M_{\odot}$,
while the mass of a gas particle is $4.3 \times 10^{6} M_{\odot}$
We assign to the  halo a rigid rotation with an angular speed 
$\lambda~=~0.08$.

\begin{figure}
\psfig{file=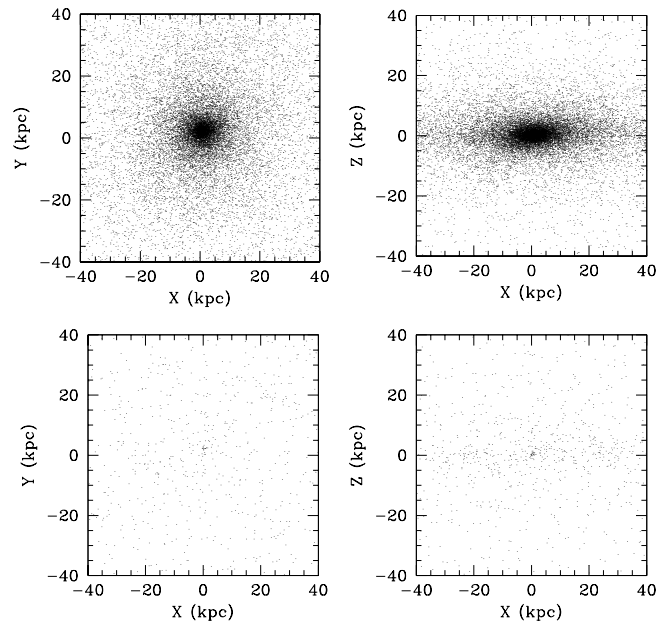,width=9truecm}
\caption{Distribution in the x-y and x-z planes of star particles 
({\it top panels}) and of gas particles ({\it bottom panels}) 
in the innermost 40~kpc, after 7~Gyrs.}
\label{discstars+gas}
\end{figure}

Due to both cooling and angular momentum, gas cools down in a thin rotating 
disk
forming stars. The distribution of star and gas particles are shown in 
Fig.~\ref{discstars+gas}.
The snapshots refer to 7~Gyr from the beginning of the simulations;
while star particles are settled into a stellar disc--like object, 
most of the gas is in the form of a hot spheroidal halo.

\begin{figure}
\psfig{file=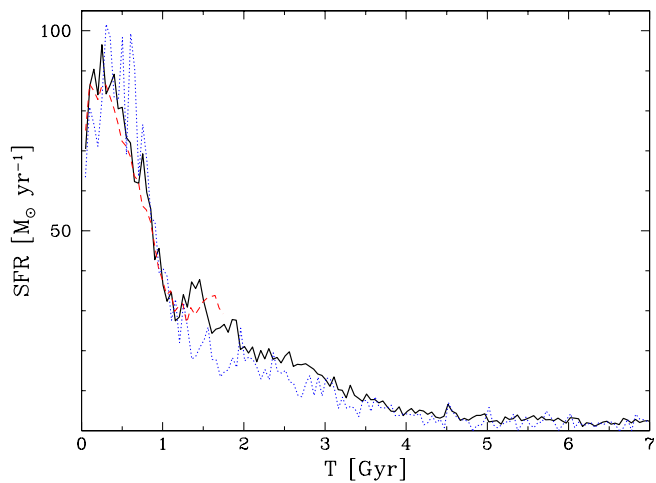,width=9truecm,angle=-90}
\caption{Global star formation history of our disc--like galaxy.
The thin dotted and dashed lines correspond to the low resolution
and very high resolution simulations, respectively.}
\label{discSFH}
\end{figure}

In Fig.~\ref{discSFH} we plot the SF history (SFH) of the overall object
(solid line).
The SFR exhibits a strong peak in the initial 0.5~Gyrs, 
then sharply declines and by the age of 4~Gyrs settles to a rather 
low constant value of a few \Msol/yr.
Correspondingly, in the final Gyrs of the simulation
most of the gas is distributed in a rather hot halo (Fig.~\ref{discstars+gas})
around the stellar disc, while little cold gas is left to fuel more active
SF within the disc.

\begin{figure}
\psfig{file=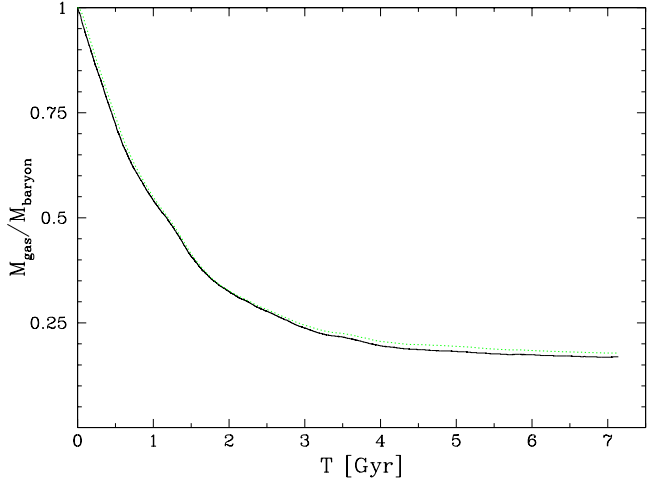,width=9truecm,angle=-90}
\caption{Evolution of the global gas fraction in our disc--like galaxy
({\it solid line}) compared to the predictions of a detailed
chemical model ({\it dotted line}). }
\label{discgasfrac}
\end{figure}

\begin{figure}
\psfig{file=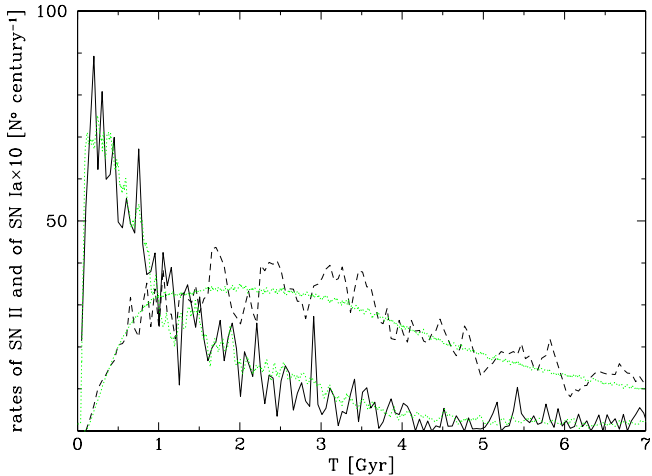,width=9truecm,angle=-90}
\caption{Evolution of the SN rates.
{\it Solid line}: type II SN; {\it dashed line}: type Ia SN (rate amplified
by a factor of 10 for the sake of clarity);
{\it dotted lines}: corresponding predictions from the chemical model.}
\label{discSN}
\end{figure}

\begin{figure}
\psfig{file=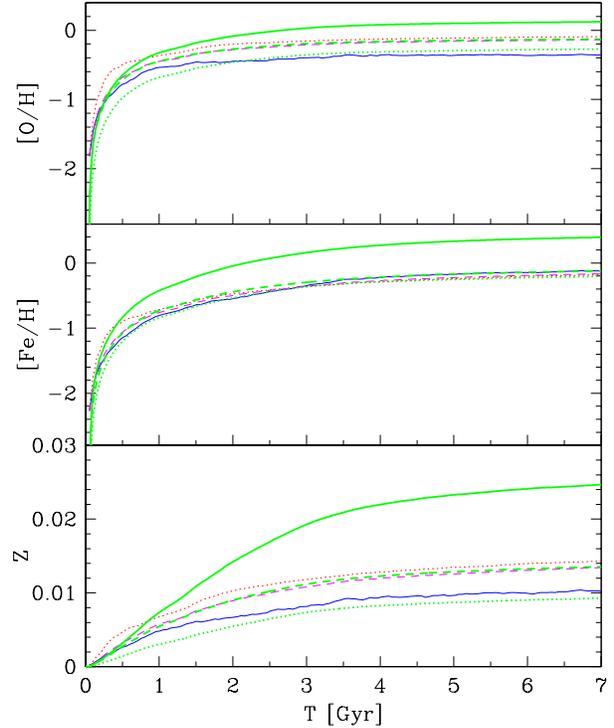,width=8truecm}
\caption{Global (gas+stars) average metal enrichment history 
({\it dashed lines})
and for gas and stars separately ({\it solid} and {\it dotted lines}, 
respectively). {\it Thin lines}: SPH simulation; {\it thick lines}:
one--zone chemical model. The global average metal enrichment compares 
very well between the two models (see text).}
\label{disczeta_tcomp}
\end{figure}

To test the behaviour of the ``chemical algorithm'' in the simulation,
we compare it with the results of a detailed one--zone chemical evolution 
model
with the same SFH: global properties like the overall gas consumption,
SN rates etc.\ should correspond. Usually, models for chemical evolution
calculate their own SFH internally after some prescribed analytical 
Schmidt--like law (e.g.\ Portinari \& Chiosi 1999). For the present purpose, 
we developed instead a version of the chemical model by PCB98 suitable to be
force--fed the SFH as an input information. The chemical
model has then been run for a closed system, just as our hydro--dynamical
galaxy is when considered globally.

Fig.~\ref{discgasfrac} shows the evolution of the global gas fraction
in our object, as SF proceeds. The dotted line is the corresponding prediction
from the chemical model, once the SFH in Fig.~\ref{discSFH} is 
imposed. The agreement is quite good.

Fig.~\ref{discSN} shows the time evolution of the SN rates for type~II and
type~Ia SN, respectively. Notice how the rate of SN~II closely traces
the SFR, as expected, while the rate of SN~Ia is much more diluted in time.
In this object with a low final SFR, the two types of SN end up with comparable
rates. The trend for the SN rates, just as for the gas consumption
in the previous figure, is in full agreement with the
predictions of the chemical model; this is a confirmation of the
good behaviour of our algorithm for gas restitution and chemical evolution
in more complex simulations than a single burst.

\begin{figure}
\psfig{file=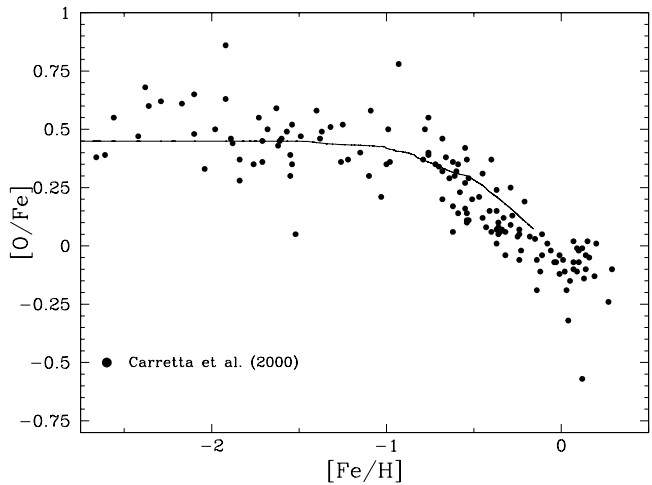,width=9truecm,angle=-90}
\caption{Evolution of the average chemical composition of the stellar 
component: [O/Fe] ratio versus metallicity}
\label{discOsuFe}
\end{figure}

\begin{figure}
\psfig{file=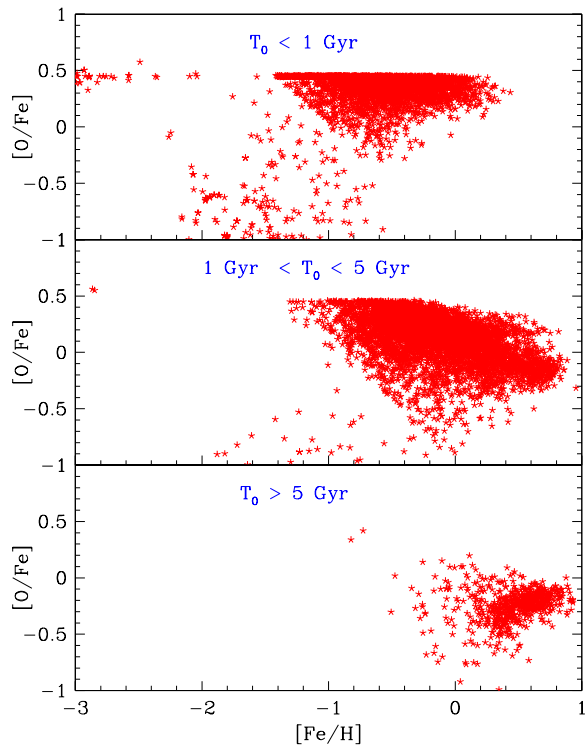,width=8truecm}
\caption{Chemical composition of stars in three different age bins
(old, intermediate and young). $T_0$ is the birth time of stars, the
overall system is 7~Gyr old.}
\label{discstarchim}
\end{figure}

\begin{figure}
\psfig{file=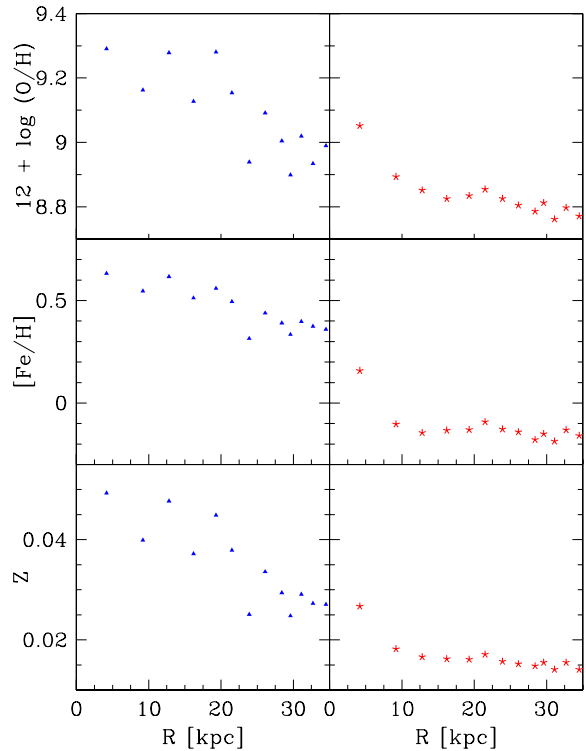,width=8truecm}
\caption{Metallicity profiles for gas ({\it left panels}) and stars 
({\it right panels})}
\label{discgradzeta}
\end{figure}

With respect to the detailed metallicity evolution of the gaseous and stellar 
components of the system, instead,
it is little meaningful to compare the hydrodynamical simulation 
and the one--zone chemical model. In fact, a basic assumption of one--zone 
models is that the system is always homogeneous in space, namely the metals
produced and ejected by stars are instantaneously mixed and spread over
the whole gas mass in the system. This is fundamentally different from the
behaviour of dynamical models of galaxies, where star formation and metal
pollution are localized. In a realistic disc--like galaxy stars pollute,
and form from, the gas in the disc region, while there will be a fraction 
of gas away from the disc which may take little or no part to SF and metal 
pollution. Qualitatively, in a dynamical simulation of a disc galaxy 
we expect stars to enrich a more limited amount of gas (that located 
in their surroundings), which gets enriched more effectively than in the 
one--zone model; this highly enriched gas is somehow ``compensated'' by the
the pristine, or almost pristine, gas residing far away from the disc.
Consequently, the stars in the disc form in an environment which is more metal
enriched than what a simple one--zone model predicts, and their metallicity
distribution is expected to be different than that derived from the chemical 
model. However, if the distribution of metals between gas and stars is
expected to be different in the dynamical simulation and in the chemical model,
the global metallicity of the {\it sum} of the
gaseous+stellar components should be comparable between the two models,
as this traces the global metal production for that particular SFH. 
Fig.~\ref{disczeta_tcomp} illustrates these expected effects: the histories 
of metal 
enrichment for gas and stars separately are evidently very different 
between the 
one--zone model and the simulation. In particular, in the hydro--dynamical
simulation more metals result locked into the stellar component, 
since the stars form in localized, very metal enriched regions;
correspondingly, the gas metallicity in the overall is lower than that 
of stars because the gas component includes also plenty of un-processed 
hot gas in the outskirts of the galaxy.
However, when one considers the global metal enrichment
(dashed lines, average of gas+stars), which is the record of the overall
metal production, the two models compare very well; minor differences 
can be imputed to the fact that the chemical model by PCB98 includes detailed
metallicity--dependent yields, while our fitting formul\ae\ in 
Table~\ref{pZtab} are based on average values of the yields.
The comparison in Fig.~\ref{disczeta_tcomp} implies that
the chemical algorithm in the SPH simulation gives an adequate description
of the metal production.

Fig.~\ref{discOsuFe} shows the evolution of the global, average [O/Fe] ratio 
for the stellar component, versus metallicity. At low metallicities stars 
display an $\alpha$--enhanced composition typical of SN II enrichment, while
at higher metallicities the [O/Fe] ratio decreases, as expected from the
additional contribution of SN~Ia to the iron production. 
Notice how the predicted evolution compares to the observational data for
stars in the Solar Neighbourhood. We remark, however, that this comparison
must be regarded only as qualitative, since the present simulation is not aimed
at reproducing the Milky Way or the Solar Neighbourhood, but only
to test the self-consistency of the chemical algorithm we implemented. 
Actually,
our object in the end resembles more the gas-consumed disc of a young S0 than
that of an Sb spiral like our own Galaxy. 

Fig.~\ref{discstarchim}
shows the chemical composition of stars in three representative age bins:
``old'' stars born within the first Gyr of the simulation, ``intermediate age''
stars, born between 1 and 5~Gyrs of the simulation, and ``young'' stars,
i.e.\ younger than 2~Gyr by the end of the simulation. 
As expected, the bulk of old stars concentrate
around high [O/Fe] ratios because little iron has yet been contributed by 
SN~Ia , while at decreasing age the bulk of stars moves to
higher metallicities and lower [O/Fe] ratios. The stars of the youngest bin
end up with super-solar metallicity and slightly under-solar [$\alpha$/Fe] 
composition.

Finally, in Fig.~\ref{discgradzeta} we show the metallicity profiles, for
gas and stars, in the innermost regions at the final age of 7~Gyr. In these
regions, where the bulk of the galaxy resides, the gas is much more metal rich
than the stars, as expected in general since the gas is an instantaneous
picture of the chemical enrichment reached at present, while stars are
a signature of the accumulated chemical history starting from the initial, 
metal--poor early epochs. Besides, in the galaxy under examination very little
gas is left in the late stages within the stellar disc, hence its chemical 
enrichment proceeds very fast.

Metallicity gradients are visible, and they are more pronounced for the gas 
component than for the stars, as expected in general from chemical evolution 
models (e.g.\ Edmunds \& Greenhow 1995).

\begin{figure}
\psfig{file=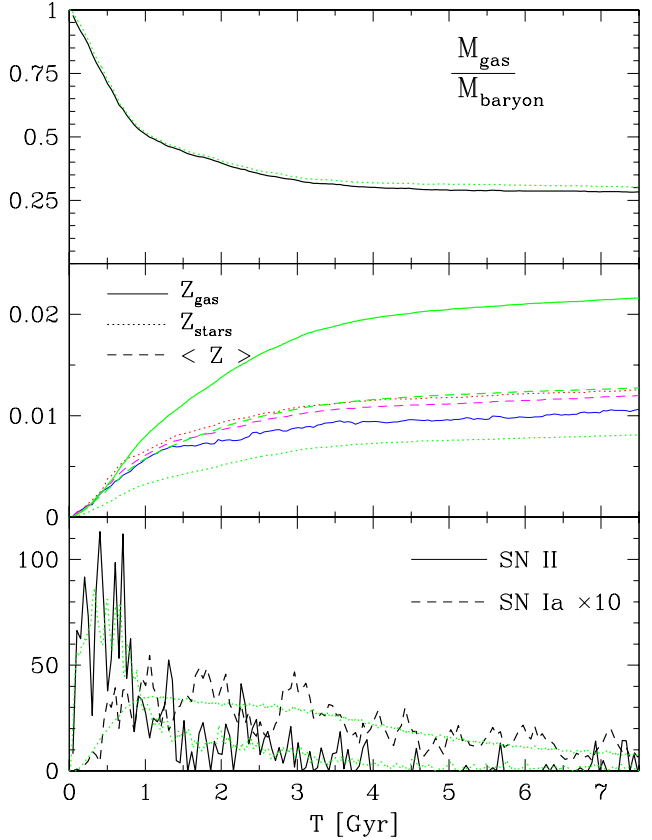,width=9truecm}
\caption{Comparison between the low resolution galaxy simulation and the
chemical one--zone model with the same SFH (the dashed line in
Fig.~\protect{\ref{discSFH}}). 
{\it Top panel}: evolution of the gas fraction; 
solid and dotted lines for simulation and chemical model respectively,
as in Fig.~\protect{\ref{discgasfrac}}.
{\it Mid panel}: metallicity evolution for gas, stars and global average;
thin and thick lines for simulation and chemical model respectively,
as in Fig.~\protect{\ref{disczeta_tcomp}}.
{\it Bottom panel}: evolution of the SN rates;
the dotted lines are the predictions of the chemical model.}
\label{lowres_comp}
\end{figure}

\subsection{Testing the effects of resolution}
A basic issue of hydrodynamical simulations of galaxy formation is
to what extent the results are affected by resolution. In principle, 
this problem might be even more crucial when a statistical algorithm
is used for SF and chemical evolution.

The resolution effect on the SFH of individual galaxies has been
explored in Lia \etal (2000) by means of {\it ad hoc} simulations
of the same individual galaxy at increasing number of particles
(2,000, 20,000 and 200,000). 
They showed that the SF recipe  converged above 20,000 particles; 
beyond that, the SFH did not change significantly by varying 
the particle number further. 

We repeat an analogous test here for our disc--like galaxy, since
our SF algorithm is different now: although we adopt the same formal
SF law as in Lia \etal (2000), this law is now implemented
with a probabilistic approach (\S~\ref{star_formation}). 
Besides, resolution tests are needed not only for the convergence of the
SFH, but also because the gas and metal restitution are treated 
statistically.

In the previous section, we discussed the self-consistency of our
chemical algorithm for a simulation with 30,000 baryonic particles, 
by comparing its results with a one--zone chemical model, where possible.
We will refer to this simulation as the ``high--resolution'' one.
Fig.~\ref{discSFH} also shows the SFH of the same galaxy modelled
with ``low resolution'' (8,000 particles, thin dotted line) and with
``very high resolution'' (200,000 particles, thin dashed line).

The low resolution case shows a much more noisy SFH, although qualitatively
the trend resembles that of the high resolution case: an initial high SFR
is mantained for $\sim$1~Gyr, after that its level declines rapidly.

Fig.~\ref{lowres_comp} illustrates the performance of the chemical algorithm
in the low resolutions case. The gas restitution (top panel), 
and consequently the metal release (mid panel), are a bit underestimated 
when compared to the predictions of the one--zone model. 
This can be imputed to the large oscillations
of the SFR; these oscillations produce even stronger noise in the gas
(and metal and SN) release, as these imply a further ``probabilistic
event'' on top of the one that induces SF. This noise is particulary evident 
in the SN rates (bottom panel). However, the effect is not large:
gas and metal restitution seems to be underestimated just by 10\% or so.
Also the SN rates in the simulation, in spite of heavy oscillations, 
on average follow the exact one--zone counterpart. 

The very high resolution simulation was followed up to 1.75~Gyr, with the sole
purpose of demonstrating that the SFH converges when the number of particles
is increased beyond 30,000. As the performances of the chemical algorithm
are very good in the high resolution simulation (see previous section),
once the convergency of the SFH is also demonstrated one can be confident
about the overall consistency of the results.

In summary, concerning resolution effects we can conclude the following.
\begin{itemize}
\item
With 8,000 particles the SFH and chemical evolution of the object
are quite noisy, and the statistical algorithm somewhat 
underestimates the gas and metal release. However, {\it for that SFH} 
the mismatch with respect to the detailed chemical model is 
by no means dramatic (within 10\%) so the gross features of chemical evolution
are already rendered.
\item
With 30,000 particles and beyond, the SFH converges and at that
level of resolution the statistical chemical algorithm also responds 
pretty well.
\end{itemize}
Concerning this latter point, a similar result was found by
Lia \etal (2000): the SFH in their simulations converged beyond 20,000
particles. Since in their case the SF algorithm was not statistical, 
we conclude that resolution effects on the SF law itself are dominant, 
while the probabilistic approach has a minor impact on the resolution limit.
Most important, we stress once more that beyond the resolution limit
for the SFH, also the statistical chemical algorithm for gas and metal 
restitution performs very well.


\section{A cluster of galaxies}

Other interesting astrophysical applications of chemo--hydrodynamical codes
lie in cosmological simulations of the formation and evolution of clusters,
to address the problem of the chemical enrichment of the ICM self-consistently.
For the sake of example, we present here one such simulation.

\begin{figure}
\psfig{file=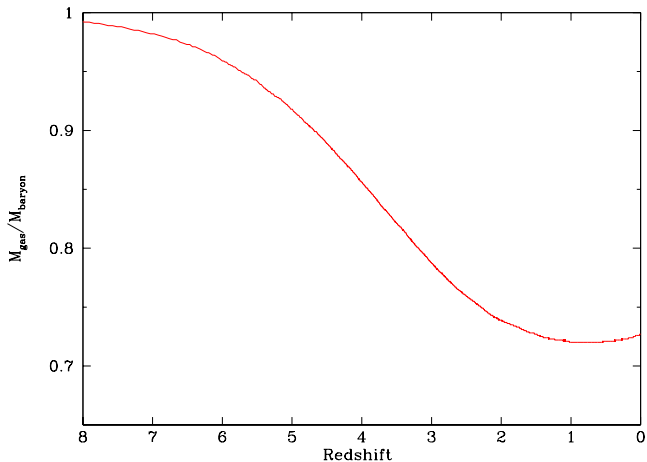,width=9truecm,angle=-90}
\caption{Evolution of the global gas fraction in the cluster}
\label{clustergasfrac}
\end{figure}

\begin{figure}
\psfig{file=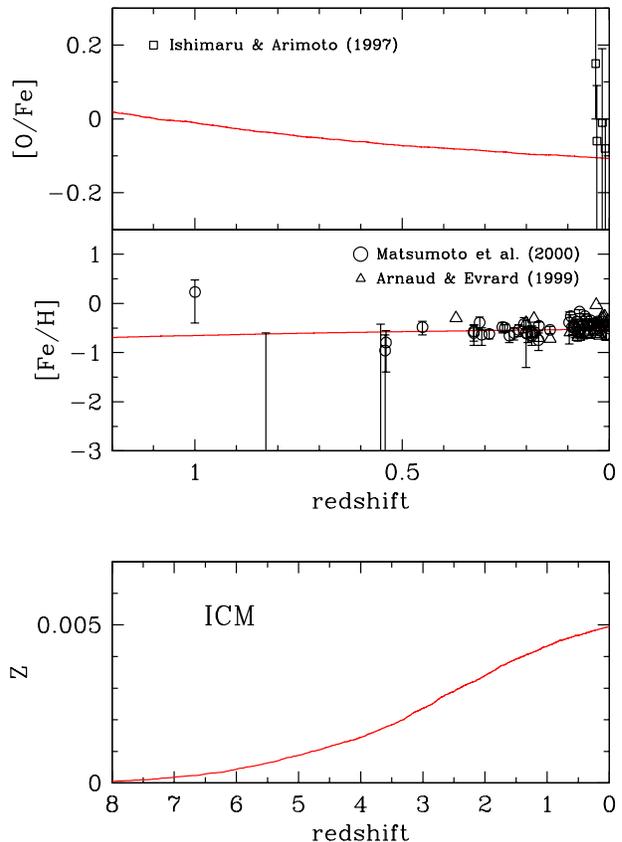,width=9truecm}
\caption{Chemical enrichment in time for the ICM}
\label{clusteramr}
\end{figure}

The initial conditions were realized by perturbing a cubic grid of particles 
with 
the displacements field made available by the {\it Cluster Comparison Project}
(http://star-www.dur.ac.uk/csf/clusdata/). The initial fluctuation spectrum was
taken to have an asymptotic spectral index, $n=1$, and shape parameter,
$\Gamma =0.25$; the cosmological parameters assumed were: mean density,
$\Omega = 1$; Hubble constant, $H_0 = 50~km s^{-1} Mpc^{-1} $; present--day
linear rms mass fluctuations in spherical top hat spheres of radius $16~Mpc$,
$\sigma_8~=~0.9$; and baryon density (in unit of the critical density),
$\Omega_b~=~0.1 $. The perturbation was centered on a cubic region of size 
$L=64~Mpc$.  See Frenk \etal (1999) for more details.
 The system was divided in two zones, an inner sphere of radius $22~Mpc$ 
which was
filled  with $36^3$ gas particles and  $36^3$ dark particles, surrounded 
by a sphere
 of radius $32 Mpc$ in which only dark matter was present. 
Initially gas and dark
 matter were placed on top of each other and were given the same velocities, 
 computed using the Zel'dovich approximation and adding the Hubble flow.
In the inner region the masses of a DM and gas particle are 
$2.1 \times 10^{10} M_{\odot}$ and $2.4 \times 10^{9} M_{\odot}$, 
respectively.  

Fig.~\ref{clustergasfrac} shows the evolution of the gas fraction 
over the total baryonic matter.
At the end of the simulation, $\sim$75\% of the baryons are in gaseous form,
while 25\% is locked into stars, compatible with observed estimates 
(the mass in gas in clusters is 2--5 times that in galaxies, 
Arnaud \etal 1992).

Fig.~\ref{clusteramr} shows the evolution of chemical abundances in the gaseous
component. The final metallicity in the ICM is $\sim$0.25 solar, 
in broad agreement with observational values.
Notice how the metalicity evolution in the gas is 
negligible at low redshifts ($z < 1$), consistent with observational data 
(Mushotzky \& Loewenstein 1997, Matsumoto et~al.\ 2000). Also the average
[O/Fe] ratio is compatible with observational data, although these are
highly uncertain (Ishimaru \& Arimoto 1997).

\begin{figure}
\psfig{file=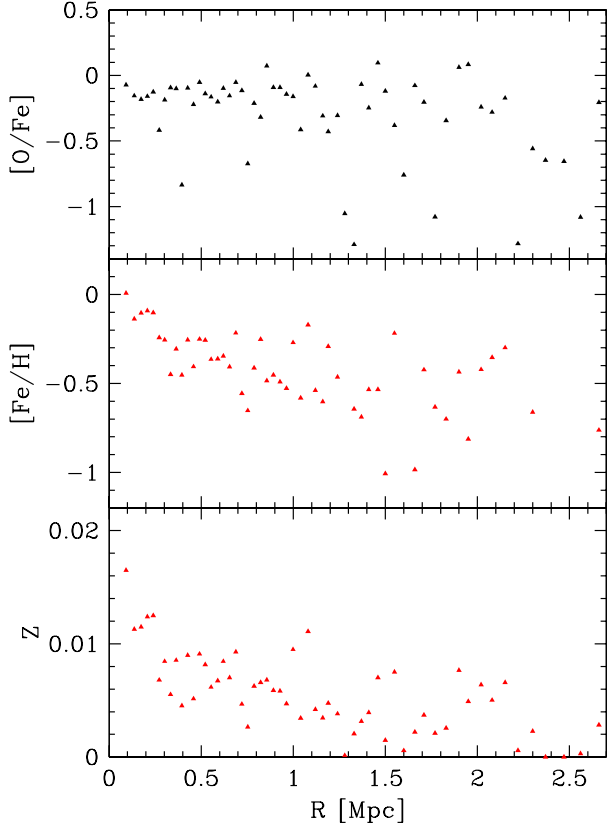,width=9truecm}
\caption{Metallicity gradients for gas in the ICM}
\label{clustergrad}
\end{figure}

\begin{figure}
\psfig{file=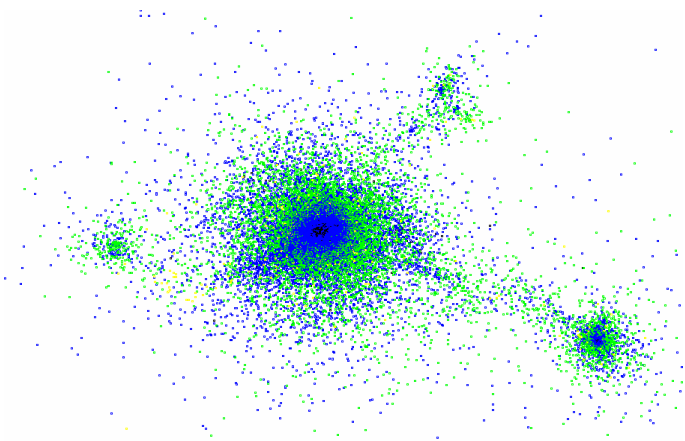,width=9truecm}
\caption{Iron distribution in the ICM: the darkest regions indicate
zones with higher iron content. See fig.\ref{clustergrad}
for comparison}
\label{ironmap}
\end{figure}

\begin{figure}
\psfig{file=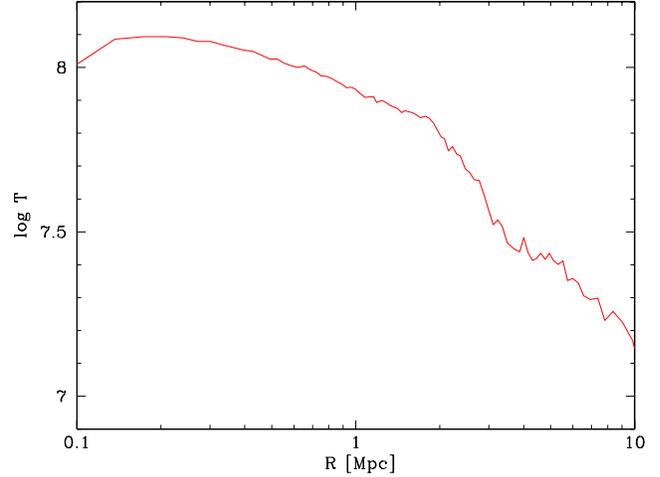,width=9truecm,angle=-90}
\caption{Radial temperature profile for the gas in the ICM}
\label{temp_clu}
\end{figure}

\begin{figure}
\psfig{file=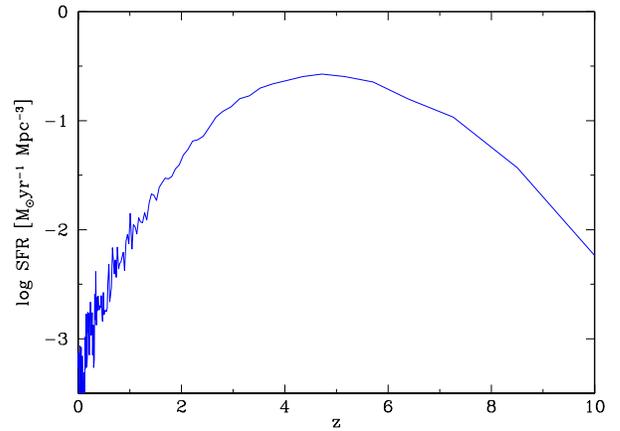,width=9truecm,angle=-90}
\caption{Global star formation history per unit volume from the whole 
simulation}
\label{cosmoSFR}
\end{figure}

Fig.~\ref{clustergrad} shows the final metallicity distribution
in the ICM. A hint of the existence of a metallicity gradient (a recent 
``hot'' issue, De Grandi \& Molendi 2001, Finoguenov \etal 2000, White 2000)
can be seen in $Z$ and in [Fe/H]. In particular,
a metallicity peak can be distinguished in the very central regions
(200--300~kpc), while further out metallicity seems to be more uniformly 
distributed, though with a large scatter. Concerning the radial behaviour
of the [$\alpha$/Fe] ratio, in our simulation there seems to be no
differential gradients for $\alpha$--elements and iron, as the [O/Fe]
ratio seems to be roughly uniform at all radii, with slightly subsolar
values, though with a large scatter.
The iron distribution is shown in Fig.~\ref{ironmap}, while the temperature
profile is plotted in Fig.~\ref{temp_clu}.

Finally, in Fig.~\ref{cosmoSFR} we show the estimated average evolution
of the SFR per Mpc$^3$ for the global simulation. Notice how the peak
on the SFR is more intense and located at earlier phases ($z \sim 5$)
with respect to what observed
in the field (e.g.\ Madau \etal 1996, Steidel \etal 1999). 
This is 
in line with expectations when looking at areas of larger density than
average; in fact, in our simulation we are dealing with a slice of
Universe where the formation of a cluster of galaxies takes place.

Higher resolution simulations are needed to assess this problem 
in the necessary detail, resolving also e.g. the gas still bound 
to individual objects from that actually expelled into the ICM, 
the role of ram pressure stripping over individual galaxies 
in the central parts of the cluster, and so forth. Such higher 
resolution simulations will be addressed in future work. In the context 
of the present paper, we presented this simulation here mostly 
for ilustration purposes, with no claims of conclusiveness.


\section{Summary and conclusions}

In this paper we developed a new algorithm for detailed calculation
of chemical evolution in SPH codes, conceived so as to be implemented
with minor computational costs into simulations with a very large number
of particles, and in parallel codes. 
As computer power and capabilities continue to improve, allowing for heavier
and heavier simulations, the opportunity arises to develop algorithms 
for the ``astrophysical calculations'' (star formation, chemical evolution
etc.) which provide increasing accuracy the
larger the number of particles involved. At the same time, as the
heaviest computational effort is 
typically devoted to the calculation of gravity forces, these ``astrophysical''
algorithms should bear as least as possible on the performances, mantaining
a low relative computational load when the number of particles increases.
Our statistical algorithm for star formation and chemical evolution, presented
in this paper, is specifically addressed to this purpose.

After testing the self--consistency of the algorithm, 
with two illustrative applications we showed how our statistical algorithm, 
implemented in a SPH code, is suitable to model 
the evolution and distribution of different chemical 
elements in various contexts. It provides a tool to follow both
the chemo--hydrodynamical evolution of individual galaxies and,
in the framework of cosmological simulations,
the chemical enrichment of the ICM/IGM as well as the global SFR and cosmic 
chemical evolution. Besides, the description of star particles as 
independent SSPs of assigned IMF, age and metallicity should make it 
straightforward to follow
spectro-photometric evolution as well, in the simulations.

Our algorithm is meant to be
easy to implement
into any SPH code.


\section*{Acknowledgements}
CL thanks C.S. Frenk for fruitful conversations
and the kind hospitality during his visit to Durham University.
LP acknowledges useful discussions with Jesper Sommer-Larsen, 
and thanks SISSA/ISAS in Trieste and the Observatory of Helsinki 
for kind hospitality on various visits. We are also grateful to Claudio Dalla
Vecchia for his help in handling the simulations. 
An anonymous referee is acknowledged for detailed remarks
on the first version of the manuscript.\\
This study was financed by
the Italian MURST (PhD and post-doc fellowships, and 
grant Cofin-9802192401) and by the Danmarks Grundforskningsfond
(TAC fellowship).


\end{document}